\documentclass[
  aps,
  prl,
  reprint,
  superscriptaddress,
  longbibliography
]{revtex4-2}

\usepackage[T1]{fontenc}
\usepackage[utf8]{inputenc}
\usepackage{lmodern}
\usepackage{amsmath,amssymb,bm}
\usepackage{hyperref}
\usepackage{graphicx}

\newcommand{\dd}{\mathrm{d}}
\newcommand{\ii}{\mathrm{i}}
\newcommand{\ee}{\mathrm{e}}
\newcommand{\vb}[1]{\bm{#1}}
\newcommand{\uv}[1]{\hat{\bm{#1}}}
\newcommand{\avg}[1]{\left\langle #1\right\rangle}
\newcommand{\abs}[1]{\left|#1\right|}

\hypersetup{colorlinks=true,
    linkcolor=blue,
    citecolor=blue,
    urlcolor=blue
}

\begin{document}

\title{Precessional modes of phonon angular momentum}

\author{Daniel~A.~Bustamante~Lopez}
\email{daniel.bustamantelopez@unifr.ch}
\affiliation{Department of Applied Physics and Science Education, Eindhoven University of Technology, 5612 AP Eindhoven, Netherlands}
\affiliation{Department of Physics, University of Fribourg, CH-1700 Fribourg, Switzerland}

\author{Dominik M. Juraschek}
\email{dominik.juraschek@unifr.ch}
\affiliation{Department of Applied Physics and Science Education, Eindhoven University of Technology, 5612 AP Eindhoven, Netherlands}
\affiliation{Department of Physics, University of Fribourg, CH-1700 Fribourg, Switzerland}
\affiliation{Paul Scherrer Institut, CH-5232 Villigen PSI, Switzerland}

\date{\today}

\begin{abstract}
Magnons are collective precessional excitations of electron spins and are ubiquitous in emerging information technologies. Here, we show that the crystal lattice can support an analogous precessional mode of phonon angular momentum. Superposition of a circularly polarized phonon and a linearly polarized phonon normal to the circular-motion plane generates a transverse angular-momentum component that rotates at their difference frequency and decays on a timescale set by the participating phonon linewidths, defining a mode with its own resonance and lifetime. Coupling between spin and phonon angular momentum hybridizes this precessional mode with a magnon near resonance. Our results establish phonon angular momentum precessional modes as lattice analogues of magnons and identify their spectroscopic and time-domain signatures.
\end{abstract}

\maketitle

\makeatletter
\let\PAMONorigaddcontentsline\addcontentsline
\renewcommand{\addcontentsline}[3]{
  \def\PAMONtoc{toc}
  \def\PAMONext{#1}
  \ifx\PAMONext\PAMONtoc
  \else
    \PAMONorigaddcontentsline{#1}{#2}{#3}
  \fi
}
\makeatother

\section{Introduction}

\begin{figure}[t]
    \centering
    \includegraphics[width=\columnwidth]{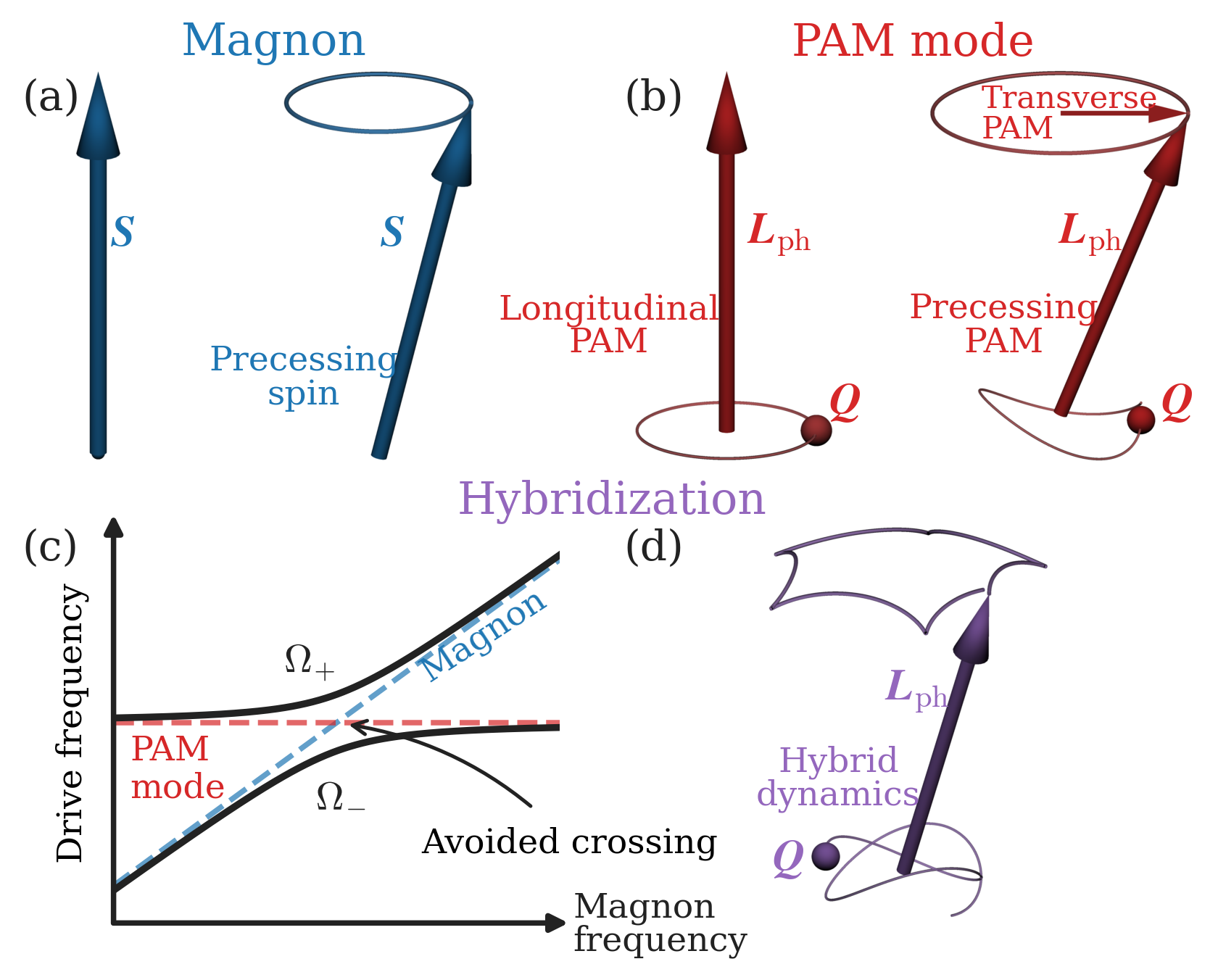}
    \caption{Coupled spin and PAM dynamics. (a) Equilibrium and precessing spin. (b) PAM precession with longitudinal and transverse components. (c) Near resonance, spin--PAM coupling produces the hybrid branches \(\Omega_{\pm}\). (d) Hybrid dynamics show reciprocal angular-momentum exchange between the magnetic and lattice subsystems.}
    \label{fig:magnonPAMHybridSchematic}
\end{figure}

Magnons are the fundamental excitations of magnetic order, representing collective precessions of electron spins, as shown in Fig.~\ref{fig:magnonPAMHybridSchematic}(a). They carry spin angular momentum, can propagate through electrically insulating magnets, and provide a basis for spin transport and wave-based information processing~\cite{Kruglyak2010,Chumak2015,Cornelissen2015,Pirro2021}. In magnetic crystals, magnons can couple to lattice vibrations and distortions~\cite{Ren2024,Mankovsky2022,Luo2025Magnophononics}, as atomic displacement or strain modulates magnetic exchange interactions and anisotropy~\cite{MaDudarev2020,Hellsvik2019,Ruckriegel2020,Mankovsky2022}. When a magnon is tuned into resonance with a phonon mode, the two hybridize into a magnon polaron~\cite{Takahashi2016,Kikkawa2016,Flebus2017,Holanda2018,Liu2021MagnonPhonon,Vaclavkova2021,Hioki2022}.

Recent work has established that lattice vibrations can themselves carry angular momentum. Circular or elliptical ionic motion produces mechanical phonon angular momentum (PAM)~\cite{ZhangNiu2014,Nakane2018,Coh2023,WangReview2024,JuraschekReview2025}, and chiral phonons carrying angular or pseudoangular momentum~\cite{Streib2021,ZhangChirality2025,Huang2026} have been observed in atomically thin and bulk crystals~\cite{Zhu2018,Ishito2023,Ueda2023}. 
Experiments have unveiled the role of PAM in angular-momentum transfer~\cite{Tauchert2022,Choi2024,Minakova2026} and transport~\cite{Ohe2024,Kim2023ChiralSSE,Nabei2026OrbitalSeebeck} processes, while experiments and theory have further shown that coherent circular excitation can generate magnetic responses, including effective magnetic fields and magnetization switching~\cite{Nova2017,Juraschek2017,Sasaki2021,Juraschek2022,Geilhufe2021,Geilhufe2023,ShabalaGeilhufe2024,Luo2023,Chaudhary2024,Basini2024,Davies2024,Yao2025,GoKim2026}. Conversely, magnetic order and magnons can couple not only to lattice displacement and strain but also to the angular momentum carried by phonons~\cite{Streib2018,Cui2023,Bonini2023CrI3,Wang2023PAM,Ma2024,Ning2024,Mrudul2025,Weissenhofer2025,Wu2025,Koller2026,Bendin2026,Mignolet2026CrI3}.
Spin couplings to lattice rotation or PAM, including terms of the form \(\vb S\cdot\vb L_{\mathrm{ph}}\), have been considered in microscopic theories of spin--phonon and spin--rotation coupling~\cite{Sheng2006,Garanin2015,HamadaMurakami2020,Kahana2024,Funato2024}, but with a focus only on PAM carried along the rotation axis of a circular phonon.

Here, we demonstrate that PAM itself can support a collective precessional mode, analogous to electron-spin precession in a magnon, as illustrated in Fig.~\ref{fig:magnonPAMHybridSchematic}(b). A superposition of a circularly polarized phonon
and another linearly polarized normal to the circular-motion plane generates a rotating transverse PAM component. In the statistical-mechanical description, this superposition is an off-diagonal coherence between the phonon modes~\cite{Simoncelli2022,Zhong2023PAM,Yi2026,Sun2026InterbandPAM}. 
This transverse component causes precession of the PAM vector about the circular phonon's rotation axis, creating a mode with its own resonance frequency and lifetime and a well-defined response to transverse perturbations. We show that spin--PAM coupling can hybridize this PAM precession with a magnon near resonance, producing the avoided crossing shown in Fig.~\ref{fig:magnonPAMHybridSchematic}(c) and the nutation dynamics shown in Fig.~\ref{fig:magnonPAMHybridSchematic}(d). Unlike conventional magnon polarons, the hybridization here is mediated by PAM rather than by the lattice displacement.

\section{Microscopic model}

\emph{Magnetic subsystem.—}We first describe the uncoupled dynamics of the magnetic and lattice degrees of freedom and then show how their coupling induces precessional modes of PAM. We consider a magnetic insulator with uniform magnetization and, for simplicity, model it as an easy-axis Heisenberg ferromagnet,
\begin{equation}
\mathcal H_{\mathrm S}
=
-\frac{J}{\hbar^2}
\sum_{\langle\alpha\beta\rangle}
\hat{\vb S}_{\alpha}\cdot\hat{\vb S}_{\beta}
-
\frac{D}{\hbar^2}
\sum_{\alpha}
\left(
\hat S_{\alpha}^{z}
\right)^2,
\label{eq:mainSpinHamiltonian}
\end{equation}
where \(\alpha\) and \(\beta\) label sites of the magnetic lattice, \(J>0\) is the nearest-neighbor ferromagnetic exchange interaction, and \(D>0\) is the easy-axis anisotropy along the crystal \(z\) direction. We work in the macrospin approximation, expanding about the equilibrium state polarized along \(+\uv z\) and retaining the spatially uniform, zone-center magnon with wave vector \(\vb q=0\), where all spins precess in phase. We denote the total spin by \(\vb S\), with equilibrium magnitude \(S_0\). In the linearized semiclassical limit of Eq.~\eqref{eq:mainSpinHamiltonian}~\cite{HolsteinPrimakoff1940}, the bare uniform-mode frequency is \(\omega_s=2Ds/\hbar\), where \(s\) is the local spin quantum number.

The dynamics are described by the Landau--Lifshitz--Gilbert (LLG) equation~\cite{Gilbert2004},
\begin{equation}
\dot{\vb S}
=
\frac{\gamma_{\mathrm{el}}}{1+\alpha_0^2}
\left[
\vb S\times\vb B_{\mathrm{eff}}
-
\frac{\alpha_0}{S_0}
\vb S\times
\left(
\vb S\times\vb B_{\mathrm{eff}}
\right)
\right],
\label{eq:mainLLG}
\end{equation}
where \(\gamma_{\mathrm{el}}<0\) is the electron gyromagnetic ratio, \(\alpha_0\) is the intrinsic Gilbert damping, and \(\vb B_{\mathrm{eff}}\) is the effective magnetic field acting on the spin. In the semiclassical description, it is obtained from the spin Hamiltonian as \(\vb B_{\mathrm{eff},\alpha}=-(1/\gamma_{\mathrm{el}})\partial\mathcal H_{\mathrm S}/\partial\vb S_\alpha\). Because the magnetic Hamiltonian is axially symmetric about \(z\), the small-amplitude uniform mode is circularly polarized in the transverse plane. Writing \(S_+=S_x+\ii S_y\), its uncoupled linearized dynamics are given by $
\left(
1+\ii\alpha_0
\right)
\dot S_+
=
-\ii\omega_s S_+$.
The corresponding uncoupled complex magnon eigenfrequency is
\(\widetilde\Omega_s^{(0)}=\omega_s/(1+\ii\alpha_0)=\Omega_s^{(0)}-\ii\kappa_s^{(0)}/2\),
with resonance frequency \(\Omega_s^{(0)}=\omega_s/(1+\alpha_0^2)\) and linewidth
\(\kappa_s^{(0)}=2\alpha_0\omega_s/(1+\alpha_0^2)\).

\emph{Lattice subsystem.—}We describe the lattice by the mass-weighted phonon coordinate
\(\vb Q=(Q_x,Q_y,Q_z)\) and its canonical momentum \(\vb P\) in the harmonic Hamiltonian
\begin{equation}
\mathcal H_{\mathrm{ph}}
=
\frac{1}{2}\vb P^2
+
\frac{1}{2}\vb Q\cdot\mathbf K\vb Q,
\qquad
\mathbf K
=
\operatorname{diag}
\left(
\omega_{\parallel}^2,
\omega_{\parallel}^2,
\omega_{\perp}^2
\right).
\label{eq:mainPhononHamiltonian}
\end{equation}
The dynamical matrix \(\mathbf K\) describes a uniaxial crystal with two degenerate in-plane phonon modes of frequency \(\omega_{\parallel}\) and an out-of-plane mode of frequency \(\omega_{\perp}\). We define the phonon angular momentum as $
\vb L_{\mathrm{ph}}
=
\vb Q\times\vb P$
~\cite{ZhangNiu2014,Nakane2018,Ruckriegel2020,Coh2023,JuraschekReview2025}. For the uncoupled harmonic lattice, Hamilton's equations give \(\vb P=\dot{\vb Q}\), so this coincides with the mechanical phonon angular momentum \(\vb Q\times\dot{\vb Q}\). The degeneracy of the two in-plane modes permits a stationary circular trajectory in the \(xy\) plane and hence a time-independent longitudinal component \(\avg{\vb L_{\mathrm{ph}}}=L_0\uv z\). In the geometry considered here, we will show that the transverse components arise from dynamical superpositions of the in-plane and out-of-plane phonon modes.

\emph{Spin--PAM coupling.—}We next introduce an effective coupling between the electron spin and the PAM,
\(\mathcal H_{\mathrm{s-ph}}=-g\,\vb S\cdot\vb L_{\mathrm{ph}}\).
The coupling strength \(g\) parametrizes the interaction and is material dependent. Microscopic mechanisms for this coupling can include spin--orbit and spin--rotation coupling to microscopic lattice rotation and PAM~\cite{HamadaMurakami2020,Geilhufe2023,Funato2024}, as well as Barnett-type conversion between phonon angular momentum and electronic spin~\cite{Qin2025}. Writing the interaction in circular components, \(A_\pm=A_x\pm\ii A_y\), separates the longitudinal and transverse couplings,
\begin{equation}
\mathcal H_{\mathrm{s-ph}}
=
-gS_zL_{\mathrm{ph},z}
-
\frac{g}{2}
\left(
S_+L_{\mathrm{ph},-}
+
S_-L_{\mathrm{ph},+}
\right).
\label{eq:mainCircularInteraction}
\end{equation}
The longitudinal interaction, familiar from earlier Raman spin--lattice and related phonon-Hall coupling theories~\cite{Sheng2006,Kagan2008}, shifts the magnon and PAM resonances through the longitudinal background values \(L_0\) and \(S_0\), whereas the transverse interaction couples their precessional dynamics and allows the two subsystems to exchange angular momentum dynamically.

The same interaction generates a torque on the spin and an equal-and-opposite torque on the lattice,
\begin{equation}
\left.
\dot{\vb S}
\right|_{\mathrm{s-ph}}
=
g\,\vb S\times\vb L_{\mathrm{ph}},
\qquad
\left.
\dot{\vb L}_{\mathrm{ph}}
\right|_{\mathrm{s-ph}}
=
-g\,\vb S\times\vb L_{\mathrm{ph}}.
\label{eq:mainMutualTorques}
\end{equation}
Accordingly, the spin--PAM interaction itself conserves total angular momentum \(\vb S+\vb L_{\mathrm{ph}}\). \(\vb S+\vb L_{\mathrm{ph}}\) need not be conserved by the full dynamics, because magnetic and lattice anisotropies, external driving, and dissipation can exert additional torques on the individual subsystems.

\emph{Collective PAM response.—}To determine how the coupling acts on the PAM, we express the harmonic phonons in terms of bosonic operators. The standard quantization of \(Q_\lambda\) and \(P_\lambda\) is given in Supplemental Material \cite{SUPP}. Within the degenerate in-plane subspace, we introduce phonons of definite circular polarization,
\(b_{\pm}=(b_x\mp\ii b_y)/\sqrt{2}\).
The longitudinal component of PAM can then be written as
\(\hat L_{\mathrm{ph},z}=\hbar(b_+^\dagger b_+-b_-^\dagger b_-)\).
Thus a population imbalance between the two oppositely circularly polarized in-plane modes yields
\(L_0=\avg{\hat L_{\mathrm{ph},z}}=\hbar(n_+-n_-)\), where
\(n_\lambda\) is the mean occupation of mode \(\lambda\). This longitudinal component is associated with circular ionic motion in the \(xy\) plane~\cite{ZhangNiu2014,Nakane2018,JuraschekReview2025}.

By contrast, the transverse PAM is carried by superpositions of the in- and out-of-plane modes and therefore evolves dynamically at their sum and difference frequencies~\cite{Zhong2023PAM,Yi2026,Sun2026InterbandPAM}. Defining \(\hat L_{\mathrm{ph},+}=\hat L_{\mathrm{ph},x}+\ii \hat L_{\mathrm{ph},y}\), one finds
\begin{equation}
\hat L_{\mathrm{ph},+}
=
A_d
\left(
b_z^\dagger b_-
-
b_+^\dagger b_z
\right)+
A_s
\left(
b_zb_-
-
b_+^\dagger b_z^\dagger
\right),
\label{eq:mainTransversePAMOperator}
\end{equation}
with $
A_{d,s}
=
\frac{\hbar}{\sqrt{2}}
\left(
\sqrt{\frac{\omega_{\parallel}}{\omega_{\perp}}}
\pm
\sqrt{\frac{\omega_{\perp}}{\omega_{\parallel}}}
\right)$.
The terms proportional to \(A_d\) transfer a phonon between the in-plane and out-of-plane modes and therefore generate difference-frequency dynamics. The terms proportional to \(A_s\) create or annihilate phonon pairs and generate sum-frequency dynamics. The transverse angular momentum is therefore carried by superpositions of different phonon modes.

We take \(\omega_{\perp}>\omega_{\parallel}\). The superposition that generates the positive-frequency difference resonance evolves under the uncoupled phonon Hamiltonian as
\(b_+^\dagger(t)b_z(t)=\ee^{-\ii(\omega_{\perp}-\omega_{\parallel})t}b_+^\dagger b_z\).
The corresponding transverse PAM oscillates at the uncoupled resonance
$\Omega_L^{(0)}
=
\omega_{\perp}-\omega_{\parallel}$.
Assuming the in-plane and out-of-plane phonons have linewidths
\(\kappa_{\parallel}\) and \(\kappa_{\perp}\), respectively, this superposition has linewidth
\(\Gamma_L=\kappa_{\parallel}+\kappa_{\perp}\), giving the uncoupled complex PAM eigenfrequency
\(\widetilde\Omega_L^{(0)}=\Omega_L^{(0)}-\ii\Gamma_L/2\).

To characterize the response of PAM to a transverse perturbation, such as that generated by spin precession through the transverse interaction in Eq.~\eqref{eq:mainCircularInteraction}, we introduce the retarded PAM susceptibility~\cite{Kubo1957},
\begin{equation}
\chi_{+-}^{R}(t)
=
-\frac{\ii}{\hbar}
\Theta(t)
\avg{
\left[
\hat L_{\mathrm{ph},+}(t),
\hat L_{\mathrm{ph},-}(0)
\right]
},
\label{eq:mainSusceptibilityDefinition}
\end{equation}
where the Heaviside function \(\Theta\) enforces causality. The susceptibility contains both the amplitude and phase of the transverse PAM response. The \(+-\) indices identify the circular response channel relevant here: the perturbation proportional to \(S_+\hat L_{\mathrm{ph},-}\), corresponding to the transverse interaction in Eq.~\eqref{eq:mainCircularInteraction}, acts on \(\hat L_{\mathrm{ph},-}\), and the induced transverse response appears in the conjugate operator \(\hat L_{\mathrm{ph},+}\). Evaluating Eq.~\eqref{eq:mainSusceptibilityDefinition} with the operator in Eq.~\eqref{eq:mainTransversePAMOperator} yields poles at the difference and sum frequencies of the participating phonons. When the magnon frequency lies close to the positive-frequency difference resonance and far from the other three poles, the response can be approximated as (see Supplemental Material \cite{SUPP})
\begin{equation}
\chi_{+-}^{R}(\omega)
\simeq
\frac{Z}{
\omega-\widetilde\Omega_L^{(0)}
},
\label{eq:mainBarePAMSusceptibility}
\end{equation}
where $
Z
=
\frac{A_d^2}{\hbar}
\left(
n_+-n_z
\right)$
is the spectral weight of the mode. It is controlled by the population difference between the two phonons participating in the superposition and can therefore be modified by temperature or nonequilibrium excitation~\cite{Hamada2018,Nova2017,Juraschek2017,Zhong2023PAM}. The difference-frequency pole of \(\chi_{+-}^{R}(\omega)\) defines the precessional PAM mode that couples to the magnon. It is a coherent excitation of the lattice angular-momentum subsystem formed from the underlying phonons, with its own resonance frequency, linewidth, and response strength.

\emph{Coupled spin--PAM dynamics and hybridization.—}Near the difference-frequency resonance, we retain from Eq.~\eqref{eq:mainTransversePAMOperator} the contribution associated with \(b_+^\dagger b_z\) and define the corresponding transverse PAM amplitude as $\ell_+\equiv\avg{\hat L_{\mathrm{ph},+}}
\simeq
-A_d\,
\avg{
b_+^\dagger b_z
}$. Linearizing around
\(\vb S=S_0\uv z\) and
\(\vb L_{\mathrm{ph}}=L_0\uv z\) yields
\begin{align}
\dot S_+
&=
-\ii
\left(
\widetilde\Omega_s^{(0)}
+
\frac{gL_0}{1+\ii\alpha_0}
\right)
S_+
+
\ii\frac{gS_0}{1+\ii\alpha_0}\ell_+,
\label{eq:mainCoupledSpin}
\\
\dot\ell_+
&=
-\ii
\left(
\widetilde\Omega_L^{(0)}
+
gS_0
\right)
\ell_+
+
\ii\frac{gZ}{2}S_+.
\label{eq:mainCoupledPAM}
\end{align}

\begin{figure}[t]
    \centering
    \includegraphics[width=\columnwidth]{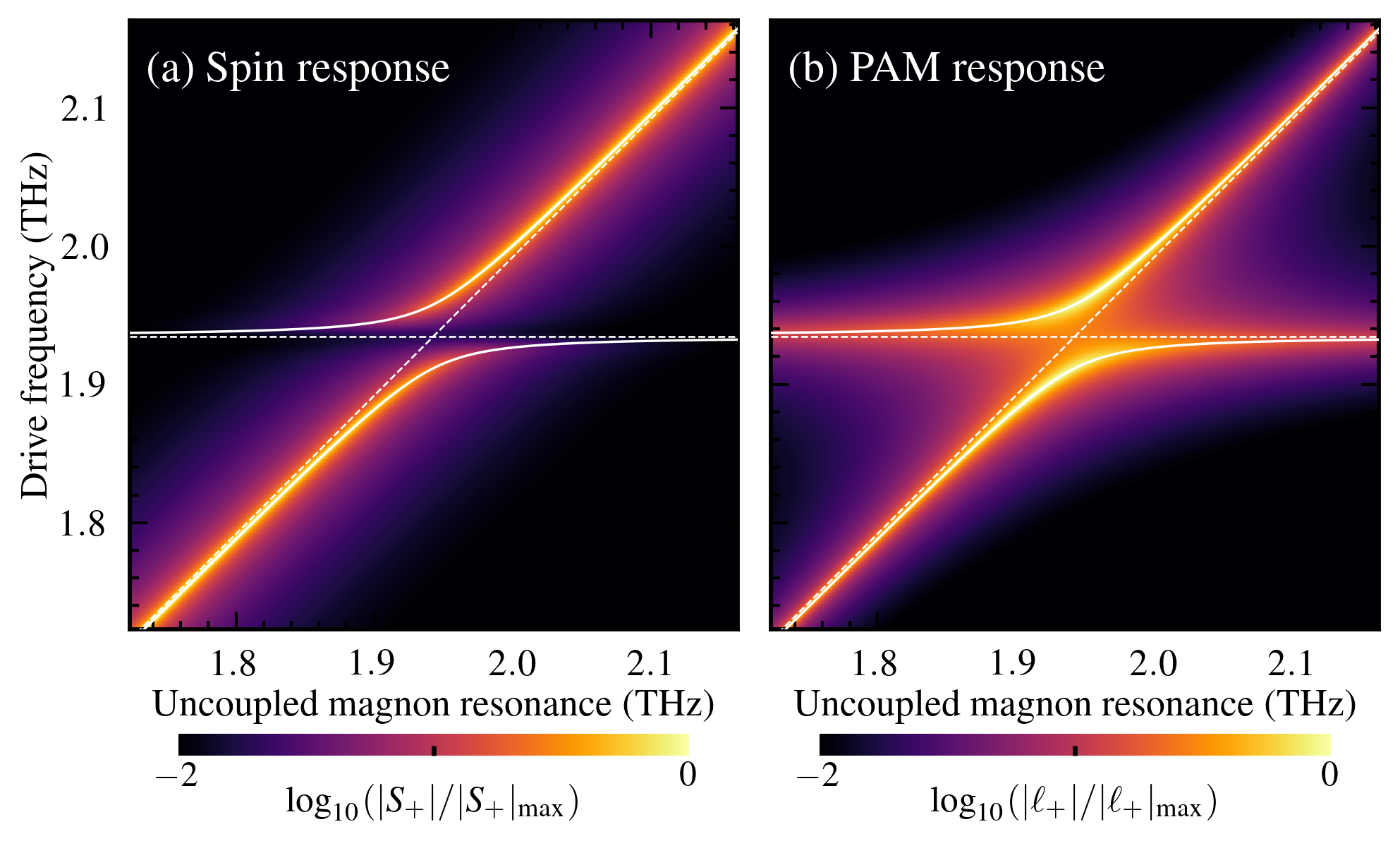}
\caption{
Driven response across the magnon--PAM hybridization region as a function of the uncoupled magnon resonance and drive frequency. Normalized circular transverse responses of (a) the spin, \( |S_+| \), and (b) the resonant PAM amplitude, \( |\ell_+| \). Dashed lines mark the shifted uncoupled resonances \(\Omega_s\) and \(\Omega_L\), while solid curves show the hybrid frequencies \(\Omega_\pm\).
}   \label{fig:magnonPAMAvoidedCrossingMap}
\end{figure}

\begin{figure*}[t]
    \centering
    \includegraphics[width=0.9\textwidth]{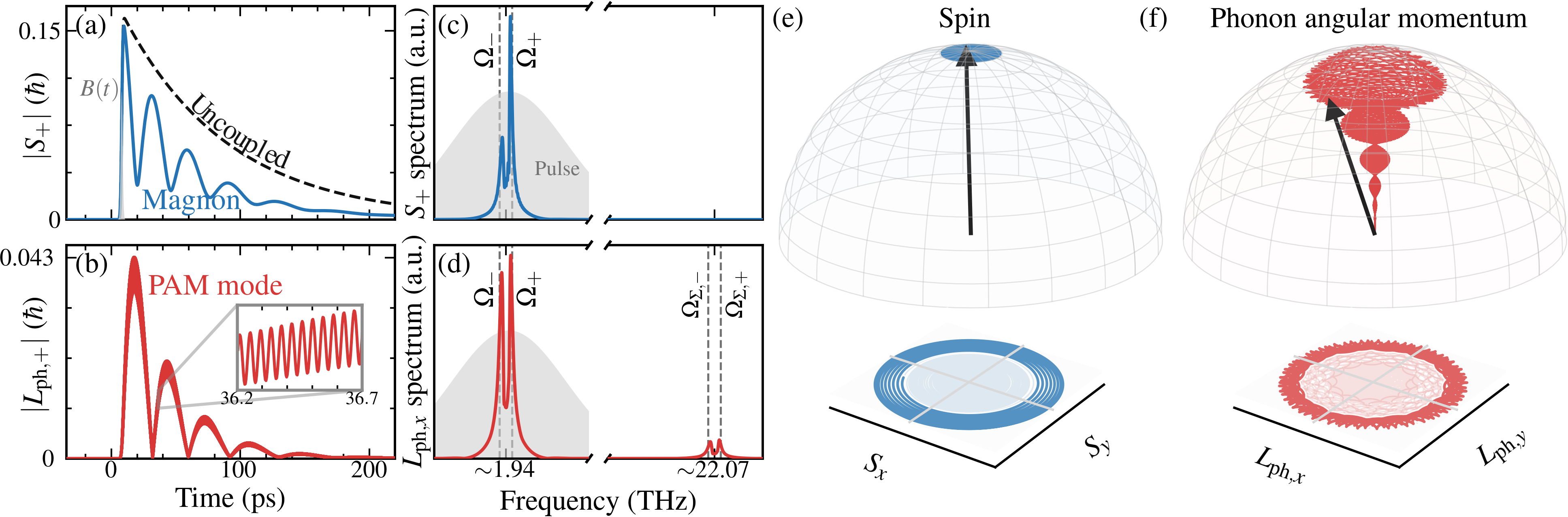}
\caption{Simulated coupled magnon--PAM dynamics following sequential circular electric- and magnetic-field excitation. (a) Time evolution of the transverse spin amplitude \( |S_{+}| \) with (blue) and without (black dashed) coupling to PAM, where the gray curve indicates the envelope of the applied magnetic-field pulse. (b) Induced transverse PAM amplitude \( |L_{\mathrm{ph},+}| \), where the inset highlights the high-frequency oscillations. (c), (d) Post-pulse spectra of \(S_{+}\) and \(L_{\mathrm{ph},x}\), respectively. The Cartesian PAM component is shown in (d) so that both the low-frequency hybrid modes and their high-frequency PAM counterparts appear at positive frequency. Dashed vertical lines mark the hybrid-mode frequencies and their high-frequency PAM counterparts, and the gray shading shows the magnetic-pulse spectrum. (e), (f) Spin and PAM vector trajectories, with the corresponding in-plane projections around the maximum precession amplitude shown below. Arrows indicate the points of maximum transverse response.}

    \label{fig:coupled-pam-spin-dynamics}
\end{figure*}

Eq.~\eqref{eq:mainCoupledSpin} describes the torque exerted by transverse PAM on the spin. Eq.~\eqref{eq:mainCoupledPAM} describes the reciprocal process in which the precessing spin generates the phonon-mode superposition that carries transverse PAM.

The longitudinal interaction shifts the uncoupled resonances. It is therefore useful to define the corresponding complex eigenfrequencies after coupling
\begin{equation}
\widetilde\Omega_s
=
\Omega_s
-
\ii\frac{\kappa_s}{2},
\qquad
\widetilde\Omega_L
=
\Omega_L
-
\ii\frac{\Gamma_L}{2},
\label{eq:mainShiftedComplexResonances}
\end{equation}
where $\Omega_s
= {(\omega_s+gL_0)}/{(1+\alpha_0^2)}$, $\kappa_s
=2\alpha_0(\omega_s+gL_0)/(1+\alpha_0^2)$, and $\Omega_L
=
\Omega_L^{(0)}+gS_0$. Taking $S_+,\ell_+\propto\ee^{-\ii\omega t}$, Eqs.~\eqref{eq:mainCoupledSpin} and \eqref{eq:mainCoupledPAM} become
\begin{equation}
\begin{pmatrix}
\omega-\widetilde\Omega_s
&
gS_0/(1+\ii\alpha_0)
\\[4pt]
gZ/2
&
\omega-\widetilde\Omega_L
\end{pmatrix}
\begin{pmatrix}
S_+
\\
\ell_+
\end{pmatrix}
=
0.
\label{eq:mainHybridMatrix}
\end{equation}
The diagonal terms describe the shifted magnon and PAM dynamics, while the off-diagonal terms describe their mutual conversion. Nontrivial solutions require the determinant to vanish, yielding two coupled complex eigenfrequencies:
\begin{equation}
\widetilde\Omega_{\pm}
=
\frac{
\widetilde\Omega_s+\widetilde\Omega_L
}{2}
\pm
\sqrt{
\frac{
\left(
\widetilde\Omega_s-\widetilde\Omega_L
\right)^2
}{4}
+
\frac{
g^2S_0Z
}{
2(1+\ii\alpha_0)
}
},
\label{eq:mainHybridPoles}
\end{equation}
where the real and imaginary parts of \(\widetilde\Omega_\pm\) determine the hybrid-mode frequencies and linewidths, respectively.

\section{Hybrid Magnon-PAM mode dynamics} 

In Fig.~\ref{fig:magnonPAMAvoidedCrossingMap}, we show the driven response of the coupled circular transverse spin and PAM amplitudes \(S_+\) and \(\ell_+\), described by Eqs.~\eqref{eq:mainCoupledSpin} and \eqref{eq:mainCoupledPAM}. A circular magnetic drive excites the magnon, and the responses in both channels exhibit an avoided crossing whose branches follow \(\Omega_\pm\). Away from resonance, the dispersive branch is magnon-like and the nearly flat branch is PAM-like, whereas near resonance both acquire mixed spin--lattice character.

We next integrate the semiclassical LLG equation for the spin \(\vb S\), Eq.~\eqref{eq:mainLLG}, together with the phonon equations of motion derived from Hamilton's equations for the coordinate and momentum \((\vb Q,\vb P)\) from Eq.~\eqref{eq:mainPhononHamiltonian}, including the spin--PAM coupling in Eq.~\eqref{eq:mainCircularInteraction} and phenomenological driving and damping, to show the coupled dynamics in real time; the explicit equations are given in Supplemental Material \cite{SUPP}. A circular electric-field pulse first generates a circularly polarized phonon carrying longitudinal PAM \(L_0\uv z\)~\cite{Nova2017,Juraschek2017,Basini2024,Davies2024}, and a time-delayed magnetic-field pulse excites the spin. In Figs.~\ref{fig:coupled-pam-spin-dynamics}(a) and \ref{fig:coupled-pam-spin-dynamics}(b), we show that the ensuing spin precession launches transverse PAM, and both modes exhibit beating at \(\abs{\Omega_+-\Omega_-}\). This directly demonstrates the coherent excitation of both hybrid-mode branches. In Figs.~\ref{fig:coupled-pam-spin-dynamics}(c) and \ref{fig:coupled-pam-spin-dynamics}(d), we show the Fourier spectra after the excitation, which resolve the same frequencies \(\Omega_-\) and \(\Omega_+\) with different weights set by the spin and PAM content of each branch. The PAM spectrum also contains the sum-frequency components \(\Omega_{\Sigma,\pm}=2(\omega_{\parallel}-gS_0)+\Omega_\pm\). These features inherit the hybrid splitting because PAM is bilinear in the phonon amplitudes. The trajectories in Figs.~\ref{fig:coupled-pam-spin-dynamics}(e) and \ref{fig:coupled-pam-spin-dynamics}(f) finally show a real-space visualization of the excitation. Instead of simple circular precession, the transverse amplitudes periodically grow and shrink, producing lobed, rosette-like projections whose repeated tips mark successive maxima of the hybrid-mode interference. The finer structure of the PAM path arises from its sum-frequency components.

Away from resonance, the coupling may equivalently be expressed as frequency-dependent field-like and damping-like torques in the LLG equation. The longitudinal PAM gives the leading static frequency shift, while the reactive and absorptive parts of the transverse PAM susceptibility renormalize the magnon frequency and linewidth. Explicit coefficients and derivations are given in Supplemental Material \cite{SUPP}.

\section{Discussion}

We have identified a precessional PAM excitation formed by a superposition of in-plane circular and out-of-plane linear phonons, whose transverse component oscillates at their difference frequency.
We stress that this PAM mode is not an additional phonon branch, but an excitation of the PAM itself with its own characteristic resonance frequency and linewidth. 
Spin precession of a magnon can couple to the transverse PAM, forming hybrid spin--PAM modes and producing an avoided crossing in the excitation spectrum when the magnon and PAM mode are near resonance.

A direct time-domain signature would be the appearance of a beating signal from the hybrid branches, \(\abs{\Omega_+-\Omega_-}\), in magnetic and lattice sensitive probes, when tuning a magnon through the PAM resonance with an external magnetic field. A well-resolved magnon--PAM hybridization requires a phonon pair with a difference frequency close to a tunable magnon resonance, sufficiently narrow phonon linewidths, and a phonon population distribution that provides both finite longitudinal PAM \(L_0\propto n_+-n_-\) and appreciable transverse spectral weight \(Z\propto n_+-n_z\). For the hybridization splitting to exceed the relevant damping scales~\cite{Liu2021MagnonPhonon,Hioki2022,Luo2025Magnophononics}, the spin--PAM coupling should further be sufficiently large. CrI$_3$ is a possible candidate to exhibit our described phenomenon~\cite{Padmanabhan2022CrI3,Bonini2023CrI3,Gou2025CrI3,Mignolet2026CrI3}, because it is a ferromagnetic insulator with a field-tunable ferromagnetic resonance and several infrared-active in-plane \(E_u\) and out-of-plane \(A_{2u}\) modes~\cite{Kim2024CrI3,Tomarchio2021CrI3} that should exhibit PAM modes at various frequencies. It has further been shown to exhibit spin--PAM coupling~\cite{Bonini2023CrI3,Ren2024}. 
Time-resolved second-harmonic generation has been used to resolve coherent infrared-active phonon motion~\cite{Taherian2025}, while time-resolved magneto-optical Kerr spectroscopy has been used to resolve spin-wave dynamics in CrI$_3$ bilayers~\cite{Zhang2020CrI3}.

The precessional dynamics identified here place PAM in close analogy with spin angular momentum, suggesting that PAM modes could be explored as lattice-based carriers for information processing and transport analogous to magnonics~\cite{ParkYang2020,Chen2021Propagating,Chen2022Diode,Suzuki2024PAMTransport}.

\bibliography{refs}

\clearpage
\onecolumngrid
\hypersetup{hidelinks}

\begin{center}
{\large\bfseries Supplemental Material for ``Precessional modes of phonon angular momentum''}
\end{center}

\renewcommand{\theHsection}{supp.\arabic{section}}
\renewcommand{\theHsubsection}{supp.\arabic{section}.\arabic{subsection}}
\renewcommand{\theHequation}{supp.\arabic{equation}}
\renewcommand{\theHfigure}{supp.\arabic{figure}}
\renewcommand{\theHtable}{supp.\arabic{table}}

\setcounter{section}{0}
\setcounter{subsection}{0}
\setcounter{equation}{0}
\setcounter{figure}{0}
\setcounter{table}{0}
\setcounter{tocdepth}{2}
\setcounter{secnumdepth}{2}

\tableofcontents

\makeatletter
\let\addcontentsline\PAMONorigaddcontentsline
\makeatother

\section{Microscopic model for coupled spin and phonon angular momentum}
\label{app:microscopic-model}

Here we construct a microscopic model for the coupling between a ferromagnetic spin mode and phonon angular momentum. The magnetic system is an easy-axis Heisenberg ferromagnet. The lattice contains two degenerate in-plane optical phonon modes and one out-of-plane optical phonon mode. We derive the uncoupled spin and phonon dynamics, the phonon-angular-momentum susceptibility, and the coupled equations used in the main text.

\subsection{Magnetic subsystem}
\label{app:magnetic-subsystem}

The magnetic Hamiltonian is
\begin{equation}
 \mathcal H_S
 =
 -\frac{J}{\hbar^2}
 \sum_{\langle\alpha\beta\rangle}
 \hat{\vb S}_{\alpha}\cdot\hat{\vb S}_{\beta}
 -
 \frac{D}{\hbar^2}
 \sum_{\alpha}
 \left(
   \hat S_{\alpha}^{z}
 \right)^2.
 \label{eq:microSpinHamiltonian}
\end{equation}
The labels \(\alpha\) and \(\beta\) identify sites of the magnetic lattice, and \(\langle\alpha\beta\rangle\) denotes a nearest-neighbor pair counted once. The operator \(\hat{\vb S}_{\alpha}\) is the local spin at site \(\alpha\), and \(J>0\) and \(D>0\) are the exchange and easy-axis anisotropy energies. The local operators satisfy \([\hat S_{\alpha}^{i},\hat S_{\beta}^{j}]=\ii\hbar\delta_{\alpha\beta}\epsilon_{ijk}\hat S_{\alpha}^{k}\) and \(\hat{\vb S}_{\alpha}^{\,2}=\hbar^2s(s+1)\), where \(s\) is the local spin quantum number and \(i,j,k\in\{x,y,z\}\).

The spin dynamics follow from the Heisenberg equation
\begin{equation}
 \frac{\dd\hat S_{\alpha}^{i}}{\dd t}
 =
 \frac{\ii}{\hbar}
 \left[
   \mathcal H_S,
   \hat S_{\alpha}^{i}
 \right].
 \label{eq:microGeneralHeisenberg}
\end{equation}
We expand about the ferromagnetic state polarized along \(+\uv z\), for which \(\langle\hat S_{\alpha}^{z}\rangle\simeq\hbar s\), and define the transverse fluctuation \(\delta\hat S_{\alpha}^{+}=\delta\hat S_{\alpha}^{x}+\ii\delta\hat S_{\alpha}^{y}\). Eq.~\eqref{eq:microGeneralHeisenberg} then gives
\begin{equation}
 \ii\hbar
 \frac{\dd}{\dd t}
 \delta\hat S_{\alpha}^{+}
 =
 2Ds\,
 \delta\hat S_{\alpha}^{+}
 +
 Js
 \sum_{\boldsymbol\delta}
 \left(
   \delta\hat S_{\alpha}^{+}
   -
   \delta\hat S_{\alpha+\boldsymbol\delta}^{+}
 \right),
 \label{eq:microLinearSpinEquation}
\end{equation}
where \(\boldsymbol\delta\) connects site \(\alpha\) to one of its nearest neighbors. The anisotropy term provides the local restoring torque, while the exchange term penalizes spatial variations of the transverse spin direction.

A spin-wave normal mode has the form \(\delta\hat S_{\alpha}^{+}(t)\propto\exp[\ii\vb q\cdot\vb R_{\alpha}-\ii\omega_{\vb q}t]\), where \(\vb R_{\alpha}\) is the position of site \(\alpha\), \(\vb q\) is a wave vector in the magnetic Brillouin zone, and \(\omega_{\vb q}\) is the bare spin-wave frequency. Substitution into Eq.~\eqref{eq:microLinearSpinEquation} gives
\begin{equation}
 \omega_{\vb q}
 =
 \frac{1}{\hbar}
 \left[
   2Ds
   +
   Js
   \sum_{\boldsymbol\delta}
   \left(
     1-
     \cos{
       \left(
         \vb q\cdot\boldsymbol\delta
       \right)
     }
   \right)
 \right].
 \label{eq:microMagnonDispersion}
\end{equation}
The magnetic mode used below is the Brillouin-zone-center mode, \(\vb q=0\). All sites then have the same transverse amplitude and phase, and the ferromagnet precesses uniformly. Within the linearized semiclassical treatment used here, its bare frequency is
\begin{equation}
 \omega_s
 =
 \omega_{\vb q=0}
 =
 \frac{2Ds}{\hbar}.
 \label{eq:microBareSpinFrequency}
\end{equation}
The exchange contribution vanishes because neighboring spins remain parallel during a uniform rotation.

The total spin is \(\hat{\vb S}_{\mathrm{tot}}=\sum_{\alpha}\hat{\vb S}_{\alpha}\), so the macrospin is \(\vb S=\langle\hat{\vb S}_{\mathrm{tot}}\rangle\) and \(S_0=N\hbar s\) in the saturated state.

Let \(S_+=S_x+\ii S_y\). Including Gilbert damping with coefficient \(\alpha_0\), the uncoupled uniform mode obeys
\begin{equation}
 \left(
   1+\ii\alpha_0
 \right)
 \dot S_+
 =
 -\ii\omega_s S_+.
 \label{eq:microDampedSpinEquation}
\end{equation}
The solution of Eq.~\eqref{eq:microDampedSpinEquation} is \(S_+(t)=S_+(0)\exp[-\ii\omega_s t/(1+\ii\alpha_0)]\). Using \(1/(1+\ii\alpha_0)=(1-\ii\alpha_0)/(1+\alpha_0^2)\), this can be written as \(S_+(t)=S_+(0)\exp[-\alpha_0\omega_s t/(1+\alpha_0^2)]\exp[-\ii\omega_s t/(1+\alpha_0^2)]\). Comparing with \(S_+(t)=S_+(0)\exp(-\kappa_s^{(0)} t/2)\exp(-\ii\Omega_s^{(0)} t)\) identifies the damped resonance frequency \(\Omega_s^{(0)}\), the magnon linewidth \(\kappa_s^{(0)}\), and the complex pole \(\widetilde\Omega_s^{(0)}=\Omega_s^{(0)}-\ii\kappa_s^{(0)}/2\):
\begin{equation}
 \widetilde\Omega_s^{(0)}
 =
 \frac{\omega_s}{1+\ii\alpha_0},
 \qquad
 \Omega_s^{(0)}
 =
 \frac{\omega_s}{1+\alpha_0^2},
 \qquad
 \kappa_s^{(0)}
 =
 \frac{2\alpha_0\omega_s}{1+\alpha_0^2}.
 \label{eq:microDampedSpinPole}
\end{equation}

\subsection{Lattice subsystem and phonon angular momentum}
\label{app:lattice-pam}

We now introduce the phonon coordinates. The mass-weighted displacement is \(\vb Q=(Q_x,Q_y,Q_z)\), and \(\vb P=(P_x,P_y,P_z)\) is its canonical momentum. The bare phonon Hamiltonian is
\begin{equation}
 \mathcal H_{\mathrm{ph}}
 =
 \frac{1}{2}\vb P^{\,2}
 +
 \frac{1}{2}
 \vb Q\cdot\mathbf K\vb Q,
 \qquad
 \mathbf K
 =
 \operatorname{diag}
 \left(
   \omega_{\parallel}^2,
   \omega_{\parallel}^2,
   \omega_{\perp}^2
 \right).
 \label{eq:microPhononHamiltonian}
\end{equation}
The \(x\) and \(y\) coordinates describe two degenerate in-plane optical phonon modes with frequency \(\omega_{\parallel}\), while \(Q_z\) describes an out-of-plane optical phonon mode with frequency \(\omega_{\perp}\). The canonical variables satisfy \([Q_i,P_j]=\ii\hbar\delta_{ij}\). Hamilton's equations give \(\dot{\vb Q}=\vb P\) and \(\dot{\vb P}=-\mathbf K\vb Q\), or equivalently \(\ddot Q_x+\omega_{\parallel}^2Q_x=0\), \(\ddot Q_y+\omega_{\parallel}^2Q_y=0\), and \(\ddot Q_z+\omega_{\perp}^2Q_z=0\).

The phonon angular momentum is
\begin{equation}
 \hat L_{\mathrm{ph},i}
 =
 \epsilon_{ijk}
 Q_jP_k.
 \label{eq:microCanonicalAngularMomentum}
\end{equation}
For the corresponding classical variables, \(\vb P=\dot{\vb Q}\), so \(\vb L_{\mathrm{ph}}=\vb Q\times\dot{\vb Q}\). A circular in-plane trajectory \(Q_x(t)=Q_0\cos(\omega_{\parallel} t)\) and \(Q_y(t)=Q_0\sin(\omega_{\parallel} t)\) has \(L_{\mathrm{ph},z}=Q_0^2\omega_{\parallel}\).

\subsection{Bosonic representation of phonon angular momentum}
\label{app:bosonic-pam}

To calculate the angular-momentum fluctuations, we quantize the three phonon modes:
\begin{equation}
 Q_\lambda
 =
 \sqrt{
   \frac{\hbar}{2\omega_\lambda}
 }
 \left(
   b_\lambda+b_\lambda^\dagger
 \right),
 \qquad
 P_\lambda
 =
 -\ii
 \sqrt{
   \frac{\hbar\omega_\lambda}{2}
 }
 \left(
   b_\lambda-b_\lambda^\dagger
 \right),
 \label{eq:microPhononQuantization}
\end{equation}
where \(\lambda\in\{x,y,z\}\), \(\omega_x=\omega_y=\omega_{\parallel}\), and \([b_\lambda,b_{\lambda'}^\dagger]=\delta_{\lambda\lambda'}\). The circular in-plane operators are
\begin{equation}
 b_+
 =
 \frac{b_x-\ii b_y}{\sqrt2},
 \qquad
 b_-
 =
 \frac{b_x+\ii b_y}{\sqrt2}.
 \label{eq:microCircularPhonons}
\end{equation}
They annihilate phonons with opposite circular polarizations in the \(xy\) plane. Substitution into Eq.~\eqref{eq:microCanonicalAngularMomentum} gives
\begin{equation}
 \hat L_{\mathrm{ph},z}
 =
 \hbar
 \left(
   b_+^\dagger b_+
   -
   b_-^\dagger b_-
 \right).
 \label{eq:microLz}
\end{equation}

For a stationary state with no coherence between the bare phonon modes, define the mean populations \(n_\lambda=\langle b_\lambda^\dagger b_\lambda\rangle\). The average phonon angular momentum is
\begin{equation}
 L_0
 =
 \avg{
   \hat L_{\mathrm{ph},z}
 }
 =
 \hbar
 \left(
   n_+-n_-
 \right).
 \label{eq:microAverageL}
\end{equation}
A population imbalance between the two circular in-plane modes therefore produces a time-independent average angular momentum along \(z\).

The transverse component \(\hat L_{\mathrm{ph},+}=\hat L_{\mathrm{ph},x}+\ii\hat L_{\mathrm{ph},y}\) connects phonons with different polarizations. Direct substitution of Eq.~\eqref{eq:microPhononQuantization} into Eq.~\eqref{eq:microCanonicalAngularMomentum} gives
\begin{equation}
 \hat L_{\mathrm{ph},+}
 =
 A_d
 \left(
   b_z^\dagger b_-
   -
   b_+^\dagger b_z
 \right)
 +
 A_s
 \left(
   b_zb_-
   -
   b_+^\dagger b_z^\dagger
 \right).
 \label{eq:microLplusFull}
\end{equation}
where
\begin{equation}
 A_d
 =
 \frac{\hbar}{\sqrt2}
 \left[
   \sqrt{
     \frac{\omega_{\parallel}}{\omega_{\perp}}
   }
   +
   \sqrt{
     \frac{\omega_{\perp}}{\omega_{\parallel}}
   }
 \right],
 \qquad
 A_s
 =
 \frac{\hbar}{\sqrt2}
 \left[
   \sqrt{
     \frac{\omega_{\parallel}}{\omega_{\perp}}
   }
   -
   \sqrt{
     \frac{\omega_{\perp}}{\omega_{\parallel}}
   }
 \right].
 \label{eq:microLplusCoefficients}
\end{equation}
The terms proportional to \(A_d\) transfer one phonon between an in-plane mode and the out-of-plane mode while conserving the total phonon number. The terms proportional to \(A_s\) create or annihilate a pair of phonons.

We take \(\omega_{\perp}>\omega_{\parallel}\). The operator \(b_+^\dagger b_z\) removes one phonon from the \(z\) mode and creates one phonon in the \(b_+\) mode. Under the bare phonon Hamiltonian it evolves as \(b_+^\dagger(t)b_z(t)=\exp[-\ii(\omega_{\perp}-\omega_{\parallel})t]b_+^\dagger b_z\). The corresponding angular-momentum frequency is
\begin{equation}
 \Omega_L^{(0)}
 =
 \omega_{\perp}-\omega_{\parallel}.
 \label{eq:microAngularMomentumFrequency}
\end{equation}
The pair terms in Eq.~\eqref{eq:microLplusFull} oscillate at the sum frequency
\begin{equation}
 \Omega_\Sigma^{(0)}
 =
 \omega_{\perp}+\omega_{\parallel}.
 \label{eq:microSumFrequency}
\end{equation}
The frequencies in the angular-momentum response are therefore sums and differences of phonon frequencies.

\subsection{PAM susceptibility and difference-frequency response}
\label{app:pam-susceptibility}

We include phonon damping through
\begin{equation}
 \dot b_\lambda
 =
 -
 \left(
   \ii\omega_\lambda
   +
   \frac{\kappa_\lambda}{2}
 \right)
 b_\lambda
 +
 \xi_\lambda,
 \label{eq:microDampedPhononModes}
\end{equation}
where \(\kappa_\lambda\) is the linewidth of phonon mode \(\lambda\) and \(\xi_\lambda\) is the corresponding Langevin force. The complex phonon pole is \(\widetilde\Omega_\lambda=\Omega_\lambda-\ii\kappa_\lambda/2\). In this damping model, \(\Omega_\lambda=\omega_\lambda\). The difference-frequency resonance is \(\Omega_L^{(0)}=\omega_{\perp}-\omega_{\parallel}\), and the coherence obeys
\begin{equation}
 \frac{\dd}{\dd t}
 \left(
   b_+^\dagger b_z
 \right)
 =
 -
 \left(
   \ii\Omega_L^{(0)}
   +
   \frac{\Gamma_L}{2}
 \right)
 b_+^\dagger b_z
 +
 \xi_L,
 \qquad
 \Gamma_L
 =
 \kappa_{\parallel}+\kappa_{\perp}.
 \label{eq:microDampedCoherence}
\end{equation}
Its complex pole is \(\widetilde\Omega_L^{(0)}=\Omega_L^{(0)}-\ii\Gamma_L/2\). The sum-frequency terms have \(\Omega_\Sigma^{(0)}=\omega_{\perp}+\omega_{\parallel}\) and, for independent damping of the two phonon modes, \(\Gamma_\Sigma=\kappa_{\parallel}+\kappa_{\perp}\).

The transverse retarded susceptibility is
\begin{equation}
 \chi_{+-}^{R}(t)
 =
 -\frac{\ii}{\hbar}
 \Theta(t)
 \avg{
   \left[
     \hat L_{\mathrm{ph},+}(t),
     \hat L_{\mathrm{ph},-}(0)
   \right]
 }.
 \label{eq:microSusceptibilityDefinition}
\end{equation}
To evaluate it, we use the time dependence of the four bilinears in Eq.~\eqref{eq:microLplusFull}:
\begin{equation}
\begin{aligned}
 b_+^\dagger(t)b_z(t)
 &=
 \ee^{-\ii\Omega_L^{(0)}t}
 b_+^\dagger b_z,
 &\qquad
 b_z^\dagger(t)b_-(t)
 &=
 \ee^{+\ii\Omega_L^{(0)}t}
 b_z^\dagger b_-,
 \\
 b_z(t)b_-(t)
 &=
 \ee^{-\ii\Omega_\Sigma^{(0)} t}
 b_zb_-,
 &
 b_+^\dagger(t)b_z^\dagger(t)
 &=
 \ee^{+\ii\Omega_\Sigma^{(0)} t}
 b_+^\dagger b_z^\dagger.
\end{aligned}
\label{eq:microBilinearTimeDependence}
\end{equation}
Define the phonon number operator \(\hat n_\lambda=b_\lambda^\dagger b_\lambda\). Using \(b_\lambda b_\lambda^\dagger=1+\hat n_\lambda\) and the fact that operators belonging to different modes commute, one representative commutator is
\begin{equation}
\begin{aligned}
 \left[
   b_+^\dagger b_z,
   b_z^\dagger b_+
 \right]
 &=
 b_+^\dagger
 \left(
   1+\hat n_z
 \right)
 b_+
 -
 b_z^\dagger
 \left(
   1+\hat n_+
 \right)
 b_z
 \\
 &=
 \hat n_+-\hat n_z.
\end{aligned}
\label{eq:microRepresentativeCommutator}
\end{equation}
More generally, for two distinct modes \(a\) and \(b\),
\begin{equation}
 \left[
   b_a^\dagger b_b,
   b_b^\dagger b_a
 \right]
 =
 \hat n_a-\hat n_b,
 \qquad
 \left[
   b_a b_b,
   b_b^\dagger b_a^\dagger
 \right]
 =
 1+\hat n_a+\hat n_b.
 \label{eq:microGeneralBilinearCommutators}
\end{equation}
For a stationary state whose density matrix is diagonal in the phonon-number basis, the remaining diagonal contributions are
\begin{equation}
\begin{aligned}
 \avg{
   \left[
     b_z^\dagger b_-,
     b_-^\dagger b_z
   \right]
 }
 &=
 n_z-n_-,
 \\
 \avg{
   \left[
     b_zb_-,
     b_-^\dagger b_z^\dagger
   \right]
 }
 &=
 1+n_z+n_-,
 \\
 \avg{
   \left[
     b_+^\dagger b_z^\dagger,
     b_zb_+
   \right]
 }
 &=
 -\left(
   1+n_z+n_+
 \right).
\end{aligned}
\label{eq:microBilinearCommutators}
\end{equation}
Eq.~\eqref{eq:microRepresentativeCommutator} gives \(\langle[b_+^\dagger b_z,b_z^\dagger b_+]\rangle=n_+-n_z\). All remaining cross terms have zero expectation value because the density matrix is diagonal in the phonon-number basis. We include damping by multiplying the difference-frequency correlations by \(\exp(-\Gamma_Lt/2)\) and the sum-frequency correlations by \(\exp(-\Gamma_\Sigma t/2)\). Using
\begin{equation}
 -\ii
 \int_0^\infty
 \dd t\,
 \ee^{\ii(\omega-\Omega)t-\Gamma t/2}
 =
 \frac{1}
 {\omega-\Omega+\ii\Gamma/2},
 \label{eq:microRetardedFourierIntegral}
\end{equation}
Eqs.~\eqref{eq:microLplusFull} and \eqref{eq:microSusceptibilityDefinition} give
\begin{align}
 \chi_{+-}^{R}(\omega)
 &=
 \frac{A_d^{\,2}}{\hbar}
 \left[
   \frac{
     n_+-n_z
   }{
     \omega-\Omega_L^{(0)}+\ii\Gamma_L/2
   }
   +
   \frac{
     n_z-n_-
   }{
     \omega+\Omega_L^{(0)}+\ii\Gamma_L/2
   }
 \right]
 \nonumber\\
 &+
 \frac{A_s^{\,2}}{\hbar}
 \left[
   \frac{
     1+n_z+n_-
   }{
     \omega-\Omega_\Sigma^{(0)}+\ii\Gamma_\Sigma/2
   }
   -
   \frac{
     1+n_z+n_+
   }{
     \omega+\Omega_\Sigma^{(0)}+\ii\Gamma_\Sigma/2
   }
 \right].
 \label{eq:microFullSusceptibility}
\end{align}
The first line contains transitions that conserve the total phonon number. The second line contains pair creation and annihilation; the constant \(1\) in the positive sum-frequency term is the vacuum contribution.

Suppose that the spin frequency lies near the positive difference-frequency pole at \(\Omega_L^{(0)}\) and far from the poles at \(\Omega_\Sigma^{(0)}\). The susceptibility can then be approximated by
\begin{equation}
 \boxed{
 \chi_{+-}^{R}(\omega)
 \simeq
 \frac{Z}{
   \omega-\Omega_L^{(0)}+\ii\Gamma_L/2
 }
 }
 \label{eq:microOnePoleSusceptibility}
\end{equation}
where
\begin{equation}
 Z
 =
 \frac{A_d^{\,2}}{\hbar}
 \left(
   n_+-n_z
 \right)
 =
 \frac{\hbar}{2}
 \left[
   \sqrt{
     \frac{\omega_{\parallel}}{\omega_{\perp}}
   }
   +
   \sqrt{
     \frac{\omega_{\perp}}{\omega_{\parallel}}
   }
 \right]^2
 \left(
   n_+-n_z
 \right).
 \label{eq:microResidue}
\end{equation}
In the damping model used here, \(\Omega_L^{(0)}=\omega_{\perp}-\omega_{\parallel}\). The residue is set by the population difference of the two modes connected by \(\hat L_{\mathrm{ph},+}\). We take \(n_+>n_z\), so \(Z>0\). The one-pole response used below and in the main text is therefore specified by \(\Omega_L^{(0)}\), \(\Gamma_L\), and \(Z\).

In thermal equilibrium, \(n_\lambda=n_{\mathrm B}(\omega_\lambda,T)=\{\exp[\hbar\omega_\lambda/(k_{\mathrm B}T)]-1\}^{-1}\). Since the two circular in-plane modes are degenerate, \(n_+=n_-\) and \(L_0=0\). The direct first-order spin shift is then absent. At finite temperature, \(n_+\neq n_z\) when \(\omega_{\parallel}\neq\omega_{\perp}\), so the difference-frequency susceptibility remains finite. At zero temperature, \(n_+=n_z=0\), so the number-conserving difference-frequency pole has zero residue, while the sum-frequency vacuum contribution remains.

The susceptibility in Eq.~\eqref{eq:microFullSusceptibility} describes a stationary phonon state whose density matrix is diagonal in the occupations of the bare modes. Its positive difference-frequency pole is carried by the freely evolving coherence \(b_+^\dagger b_z\), which oscillates at \(\Omega_L^{(0)}=\omega_{\perp}-\omega_{\parallel}\) and decays with \(\Gamma_L=\kappa_{\parallel}+\kappa_{\perp}\). Nonequilibrium excitation can modify this stationary response by changing the phonon populations, or it can generate an additional response when the \(b_+\) mode is driven coherently.

\subsection{Nonequilibrium phonon populations and coherent driving}
\label{app:nonequilibrium-pam}

Consider first an incoherent pump that creates \(b_+\) phonons through independent excitation events with no fixed phase relation between successive events. A simple population equation is
\begin{equation}
 \dot n_+
 =
 R_{\mathrm p}
 -
 \kappa_{\parallel}
 \left[
   n_+
   -
   n_{\mathrm B}(\omega_{\parallel},T)
 \right],
 \label{eq:microIncoherentPumpRate}
\end{equation}
where \(R_{\mathrm p}\) is the phonon injection rate. The steady population is
\begin{equation}
 n_+
 =
 n_{\mathrm B}(\omega_{\parallel},T)
 +
 N_{\mathrm p}^{(\mathrm{inc})},
 \qquad
 N_{\mathrm p}^{(\mathrm{inc})}
 =
 \frac{R_{\mathrm p}}{\kappa_{\parallel}}.
 \label{eq:microPopulationPump}
\end{equation}
Here \(N_{\mathrm p}^{(\mathrm{inc})}\) is the pump-induced increase in the average number of \(b_+\) phonons. We assume that the \(b_-\) mode remains in thermal equilibrium, \(n_-=n_{\mathrm B}(\omega_{\parallel},T)\).

If scattering and dephasing remove coherence between different number states, the density matrix is diagonal in the \(b_+\) number basis, \(\rho_+=\sum_n p_n|n\rangle\langle n|\). Since \(b_+\) changes the phonon number by one,
\begin{equation}
 \avg{b_+}
 =
 \sum_n
 p_n
 \langle n|b_+|n\rangle
 =
 0,
 \label{eq:microIncoherentMean}
\end{equation}
even though \(n_+=\langle b_+^\dagger b_+\rangle\) is larger than its thermal value. Eq.~\eqref{eq:microAverageL} then gives
\begin{equation}
 L_0
 =
 \hbar N_{\mathrm p}^{(\mathrm{inc})}.
 \label{eq:microPumpAngularMomentum}
\end{equation}
The same population increase changes the stationary residue to
\begin{equation}
 Z
 =
 \frac{A_d^{\,2}}{\hbar}
 \left[
   n_{\mathrm B}(\omega_{\parallel},T)
   +
   N_{\mathrm p}^{(\mathrm{inc})}
   -
   n_z
 \right].
 \label{eq:microIncoherentPumpResidue}
\end{equation}
The pole frequency and linewidth remain \(\Omega_L^{(0)}=\omega_{\perp}-\omega_{\parallel}\) and \(\Gamma_L=\kappa_{\parallel}+\kappa_{\perp}\).

We next consider a phase-stable circularly polarized drive at frequency \(\omega_{\mathrm d}\) with the helicity that excites the \(b_+\) mode. The driven mode obeys
\begin{equation}
 \dot b_+
 =
 -
 \left(
   \ii\omega_{\parallel}
   +
   \frac{\kappa_{\parallel}}{2}
 \right)
 b_+
 +
 \mathcal E_{\mathrm d}
 \ee^{-\ii\omega_{\mathrm d}t}
 +
 \xi_+(t),
 \label{eq:microDrivenCircularMode}
\end{equation}
where \(\mathcal E_{\mathrm d}\) is the complex drive amplitude and \(\xi_+\) is the Langevin force. Taking the expectation value and using \(\langle\xi_+\rangle=0\), the steady solution is
\begin{equation}
 \avg{
   b_+(t)
 }
 =
 \beta
 \ee^{-\ii\omega_{\mathrm d}t},
 \qquad
 \beta
 =
 \frac{
   \mathcal E_{\mathrm d}
 }{
   \kappa_{\parallel}/2
   +
   \ii
   \left(
     \omega_{\parallel}-\omega_{\mathrm d}
   \right)
 }.
 \label{eq:microCoherentAmplitude}
\end{equation}
The magnitude \(\abs{\beta}\) gives the coherent phonon amplitude, and its phase gives the phase of the phonon oscillation relative to the drive. The coherent contribution to the \(b_+\) population is
\begin{equation}
 N_{\mathrm p}^{(\mathrm{drv})}
 =
 \abs{\beta}^2.
 \label{eq:microCoherentPopulation}
\end{equation}
For a displaced thermal state with the \(b_-\) mode remaining in thermal equilibrium, \(n_+-n_-=N_{\mathrm p}^{(\mathrm{drv})}\), so \(L_0=\hbar N_{\mathrm p}^{(\mathrm{drv})}\).

We separate the coherent oscillation from the fluctuations:
\begin{equation}
 b_+(t)
 =
 \beta
 \ee^{-\ii\omega_{\mathrm d}t}
 +
 \delta b_+(t),
 \qquad
 \avg{
   \delta b_+(t)
 }
 =
 0.
 \label{eq:microCoherentDisplacement}
\end{equation}
The first term is maintained by the external drive, while \(\delta b_+\) contains the remaining quantum and thermal fluctuations.

We assume that the \(b_z\) and \(b_-\) modes have zero coherent amplitudes. Substituting Eq.~\eqref{eq:microCoherentDisplacement} into Eq.~\eqref{eq:microLplusFull} and retaining the terms proportional to \(\beta\) gives the drive-induced contribution
\begin{equation}
 \delta\hat L_{\mathrm{ph},+}^{(\mathrm{drv})}(t)
 \simeq
 -\beta^*
 \ee^{+\ii\omega_{\mathrm d}t}
 \left[
   A_db_z(t)
   +
   A_sb_z^\dagger(t)
 \right].
 \label{eq:microCoherentLplus}
\end{equation}
Since \(b_z(t)\propto\exp(-\ii\omega_{\perp}t)\), the first term varies as
\begin{equation}
 \ee^{+\ii\omega_{\mathrm d}t}b_z(t)
 \propto
 \ee^{-\ii(\omega_{\perp}-\omega_{\mathrm d})t}
 \label{eq:microCoherentDifferenceSideband}
\end{equation}
and produces a positive-frequency pole at \(\omega_{\perp}-\omega_{\mathrm d}\). Since \(b_z^\dagger(t)\propto\exp(+\ii\omega_{\perp}t)\), the second term varies as
\begin{equation}
 \ee^{+\ii\omega_{\mathrm d}t}b_z^\dagger(t)
 \propto
 \ee^{+\ii(\omega_{\perp}+\omega_{\mathrm d})t},
 \label{eq:microCoherentSumSideband}
\end{equation}
and contributes at the negative frequency \(-(\omega_{\perp}+\omega_{\mathrm d})\).

If the spin frequency lies near \(\omega_{\perp}-\omega_{\mathrm d}\) and far from the negative-frequency contribution, the rotating-wave approximation retains only
\begin{equation}
 \delta\hat L_{\mathrm{ph},+}^{(\mathrm{drv})}(t)
 \simeq
 -A_d\beta^*
 \ee^{+\ii\omega_{\mathrm d}t}
 b_z(t).
 \label{eq:microCoherentLplusRWA}
\end{equation}
The resulting pole has
\begin{equation}
 \Omega_L^{(\mathrm{drv})}
 =
 \omega_{\perp}-\omega_{\mathrm d},
 \qquad
 Z^{(\mathrm{drv})}
 =
 \frac{
   A_d^{\,2}
 }{
   \hbar
 }
 \abs{\beta}^2
 =
 \frac{
   A_d^{\,2}
 }{
   \hbar
 }
 N_{\mathrm p}^{(\mathrm{drv})}.
 \label{eq:microCoherentPoleParameters}
\end{equation}

Let \(\kappa_{\mathrm d}\) denote the full linewidth associated with fluctuations of the drive phase. The correlation of the coherent factor then decays as \(\exp(-\kappa_{\mathrm d}|t|/2)\). Since the remaining dynamical operator in Eq.~\eqref{eq:microCoherentLplusRWA} is \(b_z\), the linewidth of the drive-induced pole is
\begin{equation}
 \Gamma_L^{(\mathrm{drv})}
 =
 \kappa_{\perp}+\kappa_{\mathrm d}.
 \label{eq:microCoherentDriveLinewidth}
\end{equation}
For an ideal phase-stable drive, \(\kappa_{\mathrm d}=0\) and \(\Gamma_L^{(\mathrm{drv})}=\kappa_{\perp}\).

The resonant drive-induced response is therefore
\begin{equation}
 \chi_{+-}^{R,(\mathrm{drv})}(\omega)
 \simeq
 \frac{
   Z^{(\mathrm{drv})}
 }{
   \omega
   -
   \Omega_L^{(\mathrm{drv})}
   +
   \ii
   \Gamma_L^{(\mathrm{drv})}/2
 }.
 \label{eq:microDrivenOnePoleSusceptibility}
\end{equation}
The fluctuation term \(-A_d\delta b_+^\dagger b_z\) remains part of the stationary response. It is centered at \(\Omega_L^{(0)}=\omega_{\perp}-\omega_{\parallel}\) and has linewidth \(\Gamma_L=\kappa_{\parallel}+\kappa_{\perp}\). The stationary and drive-induced poles coincide in frequency only when \(\omega_{\mathrm d}=\omega_{\parallel}\), and their linewidths are generally different.

A coherent drive therefore has two effects. It increases the average \(b_+\) population and hence the longitudinal angular momentum \(L_0\), and it adds the pump-enhanced transverse pole in Eq.~\eqref{eq:microDrivenOnePoleSusceptibility}. An incoherent pump changes \(L_0\) and the residue of the stationary pole but does not generate this additional coherent contribution.

\subsection{Spin--PAM coupling and lattice equations of motion}
\label{app:spin-pam-coupling}

The spin and phonon angular momenta are coupled through
\begin{equation}
 \mathcal H_{\mathrm{s-ph}}
 =
 -g\,
 \hat{\vb S}_{\mathrm{tot}}\cdot\hat{\vb L}_{\mathrm{ph}}
 =
 -g\,
 \hat{\vb S}_{\mathrm{tot}}\cdot
 \left(
   \vb Q\times\vb P
 \right).
 \label{eq:microInteraction}
\end{equation}
In circular components,
\begin{equation}
 \mathcal H_{\mathrm{s-ph}}
 =
 -g\,
 \hat S_{\mathrm{tot},z}
 \hat L_{\mathrm{ph},z}
 -
 \frac{g}{2}
 \left(
   \hat S_{\mathrm{tot},+}
   \hat L_{\mathrm{ph},-}
   +
   \hat S_{\mathrm{tot},-}
   \hat L_{\mathrm{ph},+}
 \right).
 \label{eq:microInteractionCircular}
\end{equation}
The longitudinal term shifts the spin and phonon frequencies. The transverse terms couple \(S_+\) to \(L_{\mathrm{ph},-}\) and \(S_-\) to \(L_{\mathrm{ph},+}\). They therefore generate transverse phonon angular momentum from transverse spin motion and produce the corresponding torque on the spin.

Writing \(\vb L_{\mathrm{ph}}=\langle\hat{\vb L}_{\mathrm{ph}}\rangle\), the interaction torques are
\begin{equation}
 \left.
 \dot{\vb S}
 \right|_{\mathrm{s-ph}}
 =
 g\,
 \vb S\times\vb L_{\mathrm{ph}},
 \qquad
 \left.
 \dot{\vb L}_{\mathrm{ph}}
 \right|_{\mathrm{s-ph}}
 =
 -g\,
 \vb S\times\vb L_{\mathrm{ph}}.
 \label{eq:microSpinTorque}
\end{equation}

In the semiclassical coordinate equations, replacing \(\hat{\vb S}_{\mathrm{tot}}\) by \(\vb S\), the equations generated by \(\mathcal H_{\mathrm{ph}}+\mathcal H_{\mathrm{s-ph}}\) are
\begin{equation}
 \dot{\vb Q}
 =
 \vb P
 -
 g\,
 \vb S\times\vb Q,
 \qquad
 \dot{\vb P}
 =
 -\mathbf K\vb Q
 -
 g\,
 \vb S\times\vb P.
 \label{eq:microInteractingHamiltonEquations}
\end{equation}
Eliminating \(\vb P\) gives
\begin{equation}
 \ddot{\vb Q}
 +
 \mathbf K\vb Q
 +
 2g\,
 \vb S\times\dot{\vb Q}
 +
 g\,
 \dot{\vb S}\times\vb Q
 +
 g^2
 \vb S\times
 \left(
   \vb S\times\vb Q
 \right)
 =
 0.
 \label{eq:microInteractingCoordinateEquation}
\end{equation}
The equivalent Lagrangian is
\begin{equation}
 \mathcal L_{\mathrm{ph+s-ph}}
 =
 \frac{1}{2}
 \dot{\vb Q}^{\,2}
 -
 \frac{1}{2}
 \vb Q\cdot\mathbf K\vb Q
 +
 g\,
 \vb S\cdot
 \left(
   \vb Q\times\dot{\vb Q}
 \right)
 +
 \frac{g^2}{2}
 \abs{
   \vb S\times\vb Q
 }^2.
 \label{eq:microEquivalentLagrangian}
\end{equation}
The Euler--Lagrange equation obtained from Eq.~\eqref{eq:microEquivalentLagrangian} is identical to Eq.~\eqref{eq:microInteractingCoordinateEquation}. For \(g=0\), \(\vb P=\dot{\vb Q}\). For finite \(g\), the canonical momentum \(\vb P\) differs from \(\dot{\vb Q}\), so the interaction Hamiltonian must be expressed in terms of the canonical angular momentum \(\vb Q\times\vb P\).

For a static spin \(\vb S=S_0\uv z\), Eq.~\eqref{eq:microInteractingCoordinateEquation} splits the freely evolving circular in-plane phonon modes and shifts the \(b_+\) frequency from \(\omega_{\parallel}\) to \(\omega_{\parallel}-gS_0\). The stationary difference-frequency pole therefore shifts from \(\Omega_L^{(0)}\) to \(\Omega_L^{(0)}+gS_0\). For the drive-induced pole, the external source fixes the coherent oscillation frequency at \(\omega_{\mathrm d}\); the static spin changes the driven amplitude through the detuning but does not replace \(\omega_{\mathrm d}\) by the shifted bare-mode frequency in \(\Omega_L^{(\mathrm{drv})}=\omega_{\perp}-\omega_{\mathrm d}\).

\subsection{Reduced coupled spin--PAM dynamics}
\label{app:reduced-spin-pam}

The following reduction treats the stationary \(b_+^\dagger b_z\) response. The full susceptibility in Eq.~\eqref{eq:microFullSusceptibility} contains poles at both difference and sum frequencies. When the spin frequency lies near the positive stationary difference-frequency pole, Eq.~\eqref{eq:microOnePoleSusceptibility} neglects the other three poles. In this approximation, the induced transverse phonon angular momentum is
\begin{equation}
 \delta
 \avg{
   \hat L_{\mathrm{ph},+}(t)
 }
 \simeq
 \ell_+(t),
 \qquad
 \ell_+(t)
 \equiv
 -A_d
 \delta
 \avg{
   b_+^\dagger(t)b_z(t)
 }.
 \label{eq:microEllDefinition}
\end{equation}
Thus \(\ell_+\) is the part of \(\delta\langle\hat L_{\mathrm{ph},+}\rangle\) produced by the \(b_+^\dagger b_z\) coherence associated with the pole at \(+\Omega_L^{(0)}\). The approximation \(\delta\langle\hat L_{\mathrm{ph},+}\rangle\simeq\ell_+\) is valid only when the remaining poles in Eq.~\eqref{eq:microFullSusceptibility} are far from the spin frequency.

For a prescribed transverse spin motion \(S_+(t)\), Eq.~\eqref{eq:microInteractionCircular} gives the following time-dependent perturbation of the phonon Hamiltonian:
\begin{equation}
 \delta\mathcal H_{\mathrm{ph}}(t)
 =
 -\frac{g}{2}
 \left[
   S_+(t)\hat L_{\mathrm{ph},-}
   +
   S_-(t)\hat L_{\mathrm{ph},+}
 \right].
 \label{eq:microSpinSourcePerturbation}
\end{equation}
Using Eq.~\eqref{eq:microSusceptibilityDefinition} and retaining the term proportional to \(S_+(t)\hat L_{\mathrm{ph},-}\), linear response gives
\begin{equation}
 \ell_+(t)
 =
 -\frac{g}{2}
 \int_{-\infty}^{t}
 \dd t'\,
 \chi_{+-}^{R}(t-t')
 S_+(t').
 \label{eq:microEllLinearResponse}
\end{equation}
The minus sign follows from the susceptibility convention in Eq.~\eqref{eq:microSusceptibilityDefinition} together with the perturbation \(-gS_+\hat L_{\mathrm{ph},-}/2\). Before including the longitudinal spin-induced shift, Eq.~\eqref{eq:microOnePoleSusceptibility} gives
\begin{equation}
 \ell_+(\omega)
 =
 -\frac{g}{2}
 \frac{Z}
 {\omega-\Omega_L^{(0)}+\ii\Gamma_L/2}
 S_+(\omega).
 \label{eq:microBareEllResponse}
\end{equation}
The equilibrium longitudinal spin shifts the difference-frequency pole as \(\Omega_L^{(0)}\rightarrow\Omega_L^{(0)}+gS_0\), so
\begin{equation}
 \ell_+(\omega)
 =
 -\frac{g}{2}
 \frac{Z}
 {\omega-\Omega_L^{(0)}-gS_0+\ii\Gamma_L/2}
 S_+(\omega).
 \label{eq:microShiftedEllResponse}
\end{equation}
With the Fourier convention \(A(t)=\int\dd\omega\,\ee^{-\ii\omega t}A(\omega)/(2\pi)\), multiplying Eq.~\eqref{eq:microShiftedEllResponse} by its denominator and transforming back to time gives
\begin{equation}
 \dot\ell_+
 =
 -
 \left[
   \ii
   \left(
     \Omega_L^{(0)}+gS_0
   \right)
   +
   \frac{\Gamma_L}{2}
 \right]
 \ell_+
 +
 \ii
 \frac{gZ}{2}
 S_+.
 \label{eq:microCoupledL}
\end{equation}

The spin equation follows independently by linearizing the torque in Eq.~\eqref{eq:microSpinTorque} about \(\vb S=S_0\uv z\) and \(\vb L_{\mathrm{ph}}=L_0\uv z+\delta\vb L_{\mathrm{ph}}\). Since
\begin{equation}
 \left(
   \vb S\times\vb L_{\mathrm{ph}}
 \right)_+
 =
 \ii
 \left(
   S_0\ell_+-L_0S_+
 \right),
 \label{eq:microLinearizedCrossProduct}
\end{equation}
including the bare spin precession and Gilbert damping gives
\begin{equation}
 \left(
   1+\ii\alpha_0
 \right)
 \dot S_+
 =
 -\ii
 \left(
   \omega_s+gL_0
 \right)
 S_+
 +
 \ii gS_0\ell_+.
 \label{eq:microCoupledSpin}
\end{equation}
Eqs.~\eqref{eq:microCoupledSpin} and \eqref{eq:microCoupledL} form the reduced two-mode problem. In Eq.~\eqref{eq:microCoupledL}, the term \(\ii gZS_+/2\) generates the \(b_+^\dagger b_z\) coherence from the transverse spin motion. In Eq.~\eqref{eq:microCoupledSpin}, the term \(\ii gS_0\ell_+\) is the torque exerted by the resulting transverse phonon angular momentum on the spin.

Solving Eq.~\eqref{eq:microCoupledL} with the retarded boundary condition \(\ell_+(-\infty)=0\) gives
\begin{equation}
 \ell_+(t)
 =
 \ii
 \frac{gZ}{2}
 \int_{-\infty}^{t}
 \dd t'\,
 \exp
 \left\{
   -
   \left[
     \ii
     \left(
       \Omega_L^{(0)}+gS_0
     \right)
     +
     \frac{\Gamma_L}{2}
   \right]
   (t-t')
 \right\}
 S_+(t').
 \label{eq:microMemoryKernel}
\end{equation}
The exponential is the retarded lattice kernel, with oscillation frequency \(\Omega_L^{(0)}+gS_0\) and decay time \(2/\Gamma_L\).

Fourier transforming Eqs.~\eqref{eq:microCoupledSpin} and \eqref{eq:microCoupledL}, the homogeneous equations are
\begin{equation}
 \begin{pmatrix}
 (1+\ii\alpha_0)\omega-\omega_s-gL_0
 &
 gS_0
 \\
 gZ/2
 &
 \omega-\Omega_L^{(0)}-gS_0+\ii\Gamma_L/2
 \end{pmatrix}
 \begin{pmatrix}
 S_+(\omega)
 \\
 \ell_+(\omega)
 \end{pmatrix}
 =
 0.
 \label{eq:microCoupledMatrix}
\end{equation}
The pole equation is
\begin{equation}
 \left[
   (1+\ii\alpha_0)\omega
   -
   \omega_s
   -
   gL_0
 \right]
 \left[
   \omega
   -
   \Omega_L^{(0)}
   -
   gS_0
   +
   \ii\Gamma_L/2
 \right]
 -
 \frac{g^2S_0Z}{2}
 =
 0.
 \label{eq:microPoleEquation}
\end{equation}

\subsection{Spin-dominated pole and perturbative expansion}
\label{app:spin-pole-expansion}

Define
\begin{equation}
 \Delta_0
 =
 \omega_s
 -
 \left(
   1+\ii\alpha_0
 \right)
 \left(
   \Omega_L^{(0)}
   -
   \ii\Gamma_L/2
 \right).
 \label{eq:microBareDetuning}
\end{equation}
In the coupled system, the spin-dominated pole is written as \(\widetilde\Omega_s=\Omega_s-\ii\kappa_s/2\). It corresponds to the magnon resonance identified in Eq.~\eqref{eq:microDampedSpinPole}, modified by the coupling \(g\) to the phonon angular momentum. The root that continuously evolves from the uncoupled magnon pole is
\begin{equation}
 \widetilde\Omega_s
 =
 \frac{
   \omega_s+gL_0
 }{
   1+\ii\alpha_0
 }
 +
 \frac{
   g^2S_0Z
 }{
   2\Delta_0
 }
 +
 \frac{
   g^3S_0Z
   \left[
     (1+\ii\alpha_0)S_0-L_0
   \right]
 }{
   2\Delta_0^2
 }
 +
 O(g^4).
 \label{eq:microThirdOrderExpansion}
\end{equation}
For \(g=0\), this expression reduces to \(\widetilde\Omega_s=\omega_s/(1+\ii\alpha_0)\), which is the uncoupled damped magnon pole in Eq.~\eqref{eq:microDampedSpinPole}. The order-\(g\) term is the direct shift from the average phonon angular momentum. The order-\(g^2\) term is the retarded backaction from the phonon response. At order \(g^3\), the term proportional to \(S_0\) comes from the shift of the angular-momentum frequency, while the term proportional to \(-L_0\) comes from evaluating the second-order backaction at the shifted spin frequency.

Separating the real and imaginary parts of Eq.~\eqref{eq:microThirdOrderExpansion} gives the magnon frequency
\begin{equation}
 \boxed{
 \begin{aligned}
 \Omega_s
 &=
 \frac{\omega_s}{1+\alpha_0^2}
 +
 \frac{gL_0}{1+\alpha_0^2}
 \\
 &\quad
 +
 \frac{g^2S_0Z}{2}
 \frac{
   \omega_s-\Omega_L^{(0)}-\alpha_0\Gamma_L/2
 }{
   \left(
     \omega_s-\Omega_L^{(0)}-\alpha_0\Gamma_L/2
   \right)^2
   +
   \left(
     \Gamma_L/2-\alpha_0\Omega_L^{(0)}
   \right)^2
 }
 \\
 &\quad
 +
 \frac{g^3S_0Z}{2}
 \frac{
   \left(
     S_0-L_0
   \right)
   \left[
     \left(
       \omega_s-\Omega_L^{(0)}-\alpha_0\Gamma_L/2
     \right)^2
     -
     \left(
       \Gamma_L/2-\alpha_0\Omega_L^{(0)}
     \right)^2
   \right]
 }{
   \left[
     \left(
       \omega_s-\Omega_L^{(0)}-\alpha_0\Gamma_L/2
     \right)^2
     +
     \left(
       \Gamma_L/2-\alpha_0\Omega_L^{(0)}
     \right)^2
   \right]^2
 }
 \\
 &\quad
 +
 \frac{g^3\alpha_0S_0^2Z
   \left(
     \omega_s-\Omega_L^{(0)}-\alpha_0\Gamma_L/2
   \right)
   \left(
     \Gamma_L/2-\alpha_0\Omega_L^{(0)}
   \right)
 }{
   \left[
     \left(
       \omega_s-\Omega_L^{(0)}-\alpha_0\Gamma_L/2
     \right)^2
     +
     \left(
       \Gamma_L/2-\alpha_0\Omega_L^{(0)}
     \right)^2
   \right]^2
 }
 +
 O(g^4).
 \end{aligned}
 }
 \label{eq:microFrequencyExpansion}
\end{equation}
The corresponding magnon linewidth is
\begin{equation}
 \boxed{
 \begin{aligned}
 \kappa_s
 &=
 \frac{2\alpha_0\omega_s}{1+\alpha_0^2}
 +
 \frac{2\alpha_0gL_0}{1+\alpha_0^2}
 \\
 &\quad
 +
 g^2S_0Z
 \frac{
   \Gamma_L/2-\alpha_0\Omega_L^{(0)}
 }{
   \left(
     \omega_s-\Omega_L^{(0)}-\alpha_0\Gamma_L/2
   \right)^2
   +
   \left(
     \Gamma_L/2-\alpha_0\Omega_L^{(0)}
   \right)^2
 }
 \\
 &\quad
 +
 g^3S_0Z
 \frac{
   2
   \left(
     S_0-L_0
   \right)
   \left(
     \omega_s-\Omega_L^{(0)}-\alpha_0\Gamma_L/2
   \right)
   \left(
     \Gamma_L/2-\alpha_0\Omega_L^{(0)}
   \right)
 }{
   \left[
     \left(
       \omega_s-\Omega_L^{(0)}-\alpha_0\Gamma_L/2
     \right)^2
     +
     \left(
       \Gamma_L/2-\alpha_0\Omega_L^{(0)}
     \right)^2
   \right]^2
 }
 \\
 &\quad
 -
 g^3\alpha_0S_0^2Z
 \frac{
   \left[
     \left(
       \omega_s-\Omega_L^{(0)}-\alpha_0\Gamma_L/2
     \right)^2
     -
     \left(
       \Gamma_L/2-\alpha_0\Omega_L^{(0)}
     \right)^2
   \right]
 }{
   \left[
     \left(
       \omega_s-\Omega_L^{(0)}-\alpha_0\Gamma_L/2
     \right)^2
     +
     \left(
       \Gamma_L/2-\alpha_0\Omega_L^{(0)}
     \right)^2
   \right]^2
 }
 +
 O(g^4).
 \end{aligned}
 }
 \label{eq:microLinewidthExpansion}
\end{equation}
Eqs.~\eqref{eq:microFrequencyExpansion} and \eqref{eq:microLinewidthExpansion} are the real and imaginary parts of the same spin-dominated pole. The quadratic terms describe the reactive and dissipative parts of the retarded phonon backaction. At cubic order, the terms proportional to \(S_0\) arise from the coupling-induced shift of the angular-momentum pole, whereas those proportional to \(L_0\) arise from evaluating the backaction at the shifted spin frequency.

The derivation treats \(L_0\), \(Z\), and \(\Gamma_L\) as fixed properties of the phonon state. This is appropriate when the external drive and the lattice environment determine the phonon populations and damping, while the spin coupling is sufficiently weak that it does not change them at the order retained.

\section{Field-like and damping-like PAM torques in the LLG equation}
\label{app:transfer-coefficients}

The coupling to phonon angular momentum modifies both the precessional and dissipative parts of the spin dynamics. We describe these corrections in terms of field-like and damping-like torques and then determine their coefficients from the phonon-angular-momentum response derived above.

\subsection{LLG equation and PAM torque decomposition}
\label{app:llg-pam-torque}

The spin dynamics are written in the explicit form of the Landau--Lifshitz--Gilbert equation,
\begin{equation}
 \dot{\vb S}
 =
 \frac{\gamma_{\mathrm{el}}}{1+\alpha_0^2}
 \left[
   \vb S\times\vb B_{\mathrm{eff}}
   -
   \frac{\alpha_0}{S_0}
   \vb S\times
   \left(
     \vb S\times\vb B_{\mathrm{eff}}
   \right)
 \right]
 +
 \vb{\tau}_{\mathrm{ph}},
\label{eq:coeffExplicitLLG}
\end{equation}
Here \(\gamma_{\mathrm{el}}<0\) is the electron gyromagnetic ratio, \(\vb B_{\mathrm{eff}}\) is the effective magnetic field generated by the spin Hamiltonian, and \(\vb{\tau}_{\mathrm{ph}}\) contains the additional contribution produced by the coupling to phonon angular momentum. In the semiclassical description, the effective field acting on site \(\alpha\) is
\begin{equation}
 \vb B_{\mathrm{eff},\alpha}
 =
 -\frac{1}{\gamma_{\mathrm{el}}}
 \frac{\partial\mathcal H_S}{\partial\vb S_\alpha}.
 \label{eq:coeffEffectiveFieldDefinition}
\end{equation}
For the uniform mode, the exchange field is collinear with \(\vb S_\alpha\) and therefore produces no torque. The anisotropy term in Eq.~\eqref{eq:microSpinHamiltonian} gives
\begin{equation}
 B_{\mathrm{eff},z}
 =
 \frac{2DS_\alpha^z}{\gamma_{\mathrm{el}}\hbar^2}
 \simeq
 \frac{2Ds}{\gamma_{\mathrm{el}}\hbar}
 =
 \frac{\omega_s}{\gamma_{\mathrm{el}}},
 \qquad
 \gamma_{\mathrm{el}}B_{\mathrm{eff},z}
 =
 \omega_s.
 \label{eq:coeffEffectiveFieldFrequency}
\end{equation}
Because \(\gamma_{\mathrm{el}}<0\), the effective field associated with the state polarized along \(+\uv z\) points along \(-\uv z\), consistent with the antiparallel relation between the electron magnetic moment and its spin angular momentum.

Setting \(\vb{\tau}_{\mathrm{ph}}=0\) in Eq.~\eqref{eq:coeffExplicitLLG} gives
\begin{equation}
 \dot{\vb S}
 =
 \frac{\omega_s}{1+\alpha_0^2}
 \vb S\times\uv z
 -
 \frac{\alpha_0\omega_s}{
   \left(
     1+\alpha_0^2
   \right)
   S_0
 }
 \vb S\times
 \left(
   \vb S\times\uv z
 \right).
 \label{eq:coeffUncoupledExplicitLLG}
\end{equation}
Linearization about \(\vb S=S_0\uv z\) yields
\begin{equation}
 \dot S_+
 =
 -
 \left[
   \ii
   \frac{\omega_s}{1+\alpha_0^2}
   +
   \frac{\alpha_0\omega_s}{1+\alpha_0^2}
 \right]
 S_+,
 \label{eq:coeffUncoupledLinearized}
\end{equation}
which is equivalent to Eq.~\eqref{eq:microDampedSpinEquation} and reproduces the resonance frequency and linewidth in Eq.~\eqref{eq:microDampedSpinPole}.

We next determine the form of the coupling-induced torque. At fixed spin magnitude, the only vectors available in the axially symmetric problem are \(\vb S\), \(\uv z\), and \(\vb S\times\uv z\). The most general torque constructed from these vectors can therefore be written as
\begin{equation}
 \vb{\tau}_{\mathrm{ph}}
 =
 a\,\vb S
 +
 b\,\uv z
 +
 c\,\vb S\times\uv z,
 \label{eq:coeffGeneralTorqueExpansion}
\end{equation}
where \(a\), \(b\), and \(c\) are scalar coefficients. Conservation of the spin magnitude requires
\begin{equation}
 \vb S\cdot\vb{\tau}_{\mathrm{ph}}
 =
 aS_0^2+bS_z
 =
 0,
 \label{eq:coeffMagnitudeConstraint}
\end{equation}
so that \(b=-aS_0^2/S_z\). Eq.~\eqref{eq:coeffGeneralTorqueExpansion} becomes
\begin{equation}
 \vb{\tau}_{\mathrm{ph}}
 =
 a
 \left(
   \vb S
   -
   \frac{S_0^2}{S_z}\uv z
 \right)
 +
 c\,\vb S\times\uv z.
 \label{eq:coeffConstrainedTorque}
\end{equation}
For small transverse deviations, \(\vb S=S_0\uv z+\delta\vb S_\perp\), the first vector in Eq.~\eqref{eq:coeffConstrainedTorque} satisfies
\begin{equation}
 \vb S
 -
 \frac{S_0^2}{S_z}\uv z
 =
 \frac{1}{S_z}
 \vb S\times
 \left(
   \vb S\times\uv z
 \right)
 =
 \frac{1}{S_0}
 \vb S\times
 \left(
   \vb S\times\uv z
 \right)
 +
 O
 \left(
   \delta S_\perp^2
 \right).
 \label{eq:coeffTransverseIdentity}
\end{equation}
Thus, to linear order, the coupling-induced torque contains only two independent structures. Defining \(\Omega_{\mathrm{FL}}=c\) and \(\Gamma_{\mathrm{DL}}=-a\), we write
\begin{equation}
 \vb{\tau}_{\mathrm{ph}}
 =
 \Omega_{\mathrm{FL}}
 \vb S\times\uv z
 -
 \frac{\Gamma_{\mathrm{DL}}}{S_0}
 \vb S\times
 \left(
   \vb S\times\uv z
 \right).
 \label{eq:coeffTorqueDecomposition}
\end{equation}
The first term is perpendicular to the transverse spin and changes the precession frequency. The second term is collinear with the transverse spin and changes its decay or growth rate. Accordingly, \(\Omega_{\mathrm{FL}}\) and \(\Gamma_{\mathrm{DL}}\) are the field-like and damping-like coefficients.

Substitution of Eq.~\eqref{eq:coeffTorqueDecomposition} into Eq.~\eqref{eq:coeffExplicitLLG} gives
\begin{equation}
 \dot{\vb S}
 =
 \left[
   \frac{\omega_s}{1+\alpha_0^2}
   +
   \Omega_{\mathrm{FL}}
 \right]
 \vb S\times\uv z
 -
 \frac{1}{S_0}
 \left[
   \frac{\alpha_0\omega_s}{1+\alpha_0^2}
   +
   \Gamma_{\mathrm{DL}}
 \right]
 \vb S\times
 \left(
   \vb S\times\uv z
 \right).
 \label{eq:coeffEffectiveExplicitLLG}
\end{equation}
Using
\begin{equation}
 \left(
   \vb S\times\uv z
 \right)_+
 =
 -\ii S_+,
 \qquad
 \left[
   \vb S\times
   \left(
     \vb S\times\uv z
   \right)
 \right]_+
 =
 S_0S_+,
 \label{eq:coeffLinearizedTorqueIdentities}
\end{equation}
the linearized transverse equation is
\begin{equation}
 \dot S_+
 =
 -\ii
 \left[
   \frac{\omega_s}{1+\alpha_0^2}
   +
   \Omega_{\mathrm{FL}}
 \right]
 S_+
 -
 \left[
   \frac{\alpha_0\omega_s}{1+\alpha_0^2}
   +
   \Gamma_{\mathrm{DL}}
 \right]
 S_+.
 \label{eq:coeffEffectiveLinearized}
\end{equation}
Eq.~\eqref{eq:coeffEffectiveLinearized} shows directly that \(\Omega_{\mathrm{FL}}\) shifts the magnon frequency, whereas \(\Gamma_{\mathrm{DL}}\) changes the magnon half-linewidth.

\subsection{Microscopic torque coefficients from the PAM response}
\label{app:pam-torque-coefficients}

We now determine these coefficients for the microscopic coupling to phonon angular momentum. Before solving the Gilbert equation explicitly, the interaction in Eq.~\eqref{eq:microInteraction} contributes the torque \(g\vb S\times\vb L_{\mathrm{ph}}\), as shown in Eq.~\eqref{eq:microSpinTorque}. Its longitudinal component \(L_0\uv z\) produces the direct frequency shift in Eq.~\eqref{eq:microCoupledSpin}, whereas its transverse component is induced by the spin motion through the retarded phonon-angular-momentum susceptibility. Fourier transforming Eq.~\eqref{eq:microCoupledSpin} and using Eq.~\eqref{eq:microShiftedEllResponse} gives
\begin{equation}
 \left\{
   \left(
     1+\ii\alpha_0
   \right)
   \omega
   -
   \omega_s
   -
   gL_0
   -
   \frac{g^2S_0}{2}
   \chi_{+-}^{R}(\omega)
 \right\}
 S_+(\omega)
 =
 0.
 \label{eq:coeffMicroscopicFrequencyEquation}
\end{equation}
In the one-pole approximation, including the longitudinal spin-induced shift of the stationary angular-momentum pole,
\begin{equation}
 \chi_{+-}^{R}(\omega)
 \simeq
 \frac{Z}{
   \omega
   -
   \Omega_L^{(0)}
   -
   gS_0
   +
   \ii\Gamma_L/2
 }.
 \label{eq:coeffShiftedOnePoleSusceptibility}
\end{equation}

Because the transverse phonon response is retarded, the two torque coefficients depend on the response frequency. Dividing Eq.~\eqref{eq:coeffMicroscopicFrequencyEquation} by \(1+\ii\alpha_0\) and comparing it with the Fourier transform of Eq.~\eqref{eq:coeffEffectiveLinearized} gives
\begin{equation}
 \Omega_{\mathrm{FL}}(\omega)
 =
 \frac{1}{1+\alpha_0^2}
 \left\{
   gL_0
   +
   \frac{g^2S_0}{2}
   \left[
     \operatorname{Re}\chi_{+-}^{R}(\omega)
     +
     \alpha_0
     \operatorname{Im}\chi_{+-}^{R}(\omega)
   \right]
 \right\}
\label{eq:coeffFieldLikeSusceptibility}
\end{equation}
and
\begin{equation}
 \Gamma_{\mathrm{DL}}(\omega)
 =
 \frac{1}{1+\alpha_0^2}
 \left\{
   \alpha_0gL_0
   +
   \frac{g^2S_0}{2}
   \left[
     \alpha_0
     \operatorname{Re}\chi_{+-}^{R}(\omega)
     -
     \operatorname{Im}\chi_{+-}^{R}(\omega)
   \right]
 \right\}.
\label{eq:coeffDampingLikeSusceptibility}
\end{equation}
The terms proportional to \(\alpha_0\) arise from converting the Gilbert equation to explicit Landau--Lifshitz form. They mix the reactive and dissipative parts of the phonon response but do not represent an additional microscopic dissipation channel.

For the stationary pole, Eq.~\eqref{eq:coeffShiftedOnePoleSusceptibility} gives
\begin{equation}
 \operatorname{Re}\chi_{+-}^{R}(\omega)
 =
 Z
 \frac{\omega-\Omega_L^{(0)}-gS_0}{
   \left(\omega-\Omega_L^{(0)}-gS_0\right)^2
   +
   \left(
     \Gamma_L/2
   \right)^2
 },
 \qquad
 \operatorname{Im}\chi_{+-}^{R}(\omega)
 =
 -Z
 \frac{\Gamma_L/2}{
   \left(\omega-\Omega_L^{(0)}-gS_0\right)^2
   +
   \left(
     \Gamma_L/2
   \right)^2
 }.
 \label{eq:coeffSusceptibilityParts}
\end{equation}
The explicit coefficients are therefore
\begin{equation}
 \Omega_{\mathrm{FL}}(\omega)
 =
 \frac{1}{1+\alpha_0^2}
 \left\{
   gL_0
   +
   \frac{g^2S_0Z}{2}
   \frac{
     \omega-\Omega_L^{(0)}-gS_0
     -
     \alpha_0\Gamma_L/2
   }{
     \left(\omega-\Omega_L^{(0)}-gS_0\right)^2
     +
     \left(
       \Gamma_L/2
     \right)^2
   }
 \right\}
\label{eq:coeffFieldLikeExact}
\end{equation}
and
\begin{equation}
 \Gamma_{\mathrm{DL}}(\omega)
 =
 \frac{1}{1+\alpha_0^2}
 \left\{
   \alpha_0gL_0
   +
   \frac{g^2S_0Z}{2}
   \frac{
     \Gamma_L/2
     +
     \alpha_0
     \left(
       \omega-\Omega_L^{(0)}-gS_0
     \right)
   }{
     \left(\omega-\Omega_L^{(0)}-gS_0\right)^2
     +
     \left(
       \Gamma_L/2
     \right)^2
   }
 \right\}.
\label{eq:coeffDampingLikeExact}
\end{equation}

For a direct expansion in the coupling, we keep the real response frequency \(\omega\) fixed and expand the shifted pole denominator in powers of \(gS_0\). Eqs.~\eqref{eq:coeffFieldLikeExact} and \eqref{eq:coeffDampingLikeExact} then give
\begin{equation}
 \boxed{
 \begin{aligned}
 \Omega_{\mathrm{FL}}(\omega)
 &=
 \frac{1}{1+\alpha_0^2}
 \Bigg\{
   gL_0
   +
   \frac{g^2S_0Z}{2}
   \frac{
     \omega-\Omega_L^{(0)}-\alpha_0\Gamma_L/2
   }{
     \left(
       \omega-\Omega_L^{(0)}
     \right)^2
     +
     \left(
       \Gamma_L/2
     \right)^2
   }
 \\
 &\qquad
   +
   \frac{g^3S_0^2Z}{2}
   \frac{
     \left(
       \omega-\Omega_L^{(0)}
     \right)^2
     -
     \left(
       \Gamma_L/2
     \right)^2
     -
     \alpha_0\Gamma_L
     \left(
       \omega-\Omega_L^{(0)}
     \right)
   }{
     \left[
       \left(
         \omega-\Omega_L^{(0)}
       \right)^2
       +
       \left(
         \Gamma_L/2
       \right)^2
     \right]^2
   }
 \Bigg\}
 +
 O(g^4)
 \end{aligned}
 }
 \label{eq:coeffFieldLikeExpansion}
\end{equation}
and
\begin{equation}
 \boxed{
 \begin{aligned}
 \Gamma_{\mathrm{DL}}(\omega)
 &=
 \frac{1}{1+\alpha_0^2}
 \Bigg\{
   \alpha_0gL_0
   +
   \frac{g^2S_0Z}{2}
   \frac{
     \Gamma_L/2
     +
     \alpha_0
     \left(
       \omega-\Omega_L^{(0)}
     \right)
   }{
     \left(
       \omega-\Omega_L^{(0)}
     \right)^2
     +
     \left(
       \Gamma_L/2
     \right)^2
   }
 \\
 &\qquad
   +
   \frac{g^3S_0^2Z}{2}
   \frac{
     \Gamma_L
     \left(
       \omega-\Omega_L^{(0)}
     \right)
     +
     \alpha_0
     \left[
       \left(
         \omega-\Omega_L^{(0)}
       \right)^2
       -
       \left(
         \Gamma_L/2
       \right)^2
     \right]
   }{
     \left[
       \left(
         \omega-\Omega_L^{(0)}
       \right)^2
       +
       \left(
         \Gamma_L/2
       \right)^2
     \right]^2
   }
 \Bigg\}
 +
 O(g^4).
 \end{aligned}
 }
 \label{eq:coeffDampingLikeExpansion}
\end{equation}
The linear field-like term is generated by the average longitudinal phonon angular momentum. The quadratic terms are the dispersive and absorptive parts of the transverse phonon response. The cubic terms arise because the longitudinal spin shifts the stationary angular-momentum pole by \(gS_0\). Eqs.~\eqref{eq:coeffFieldLikeExpansion} and \eqref{eq:coeffDampingLikeExpansion} are response coefficients at fixed real \(\omega\); obtaining the pole expansion through the same order additionally requires evaluating the frequency-dependent response self-consistently, as done in Eqs.~\eqref{eq:microFrequencyExpansion} and \eqref{eq:microLinewidthExpansion}.

In the weak-coupling regime, evaluating these coefficients at the uncoupled magnon resonance gives the leading changes of the spin-dominated pole:
\begin{equation}
 \Omega_s
 -
 \frac{\omega_s}{1+\alpha_0^2}
 \simeq
 \Omega_{\mathrm{FL}}
 \left(
   \frac{\omega_s}{1+\alpha_0^2}
 \right),
 \qquad
 \frac{\kappa_s}{2}
 -
 \frac{\alpha_0\omega_s}{1+\alpha_0^2}
 \simeq
 \Gamma_{\mathrm{DL}}
 \left(
   \frac{\omega_s}{1+\alpha_0^2}
 \right).
 \label{eq:coeffPoleCorrections}
\end{equation}
More generally, the coupled frequency and linewidth follow from the complex pole equation in Eq.~\eqref{eq:microPoleEquation}, whose perturbative solution is given in Eq.~\eqref{eq:microThirdOrderExpansion}.

If the phase-stable drive-induced pole is also retained, the susceptibility entering Eqs.~\eqref{eq:coeffFieldLikeSusceptibility} and \eqref{eq:coeffDampingLikeSusceptibility} is replaced by
\begin{equation}
 \chi_{+-}^{R}(\omega)
 \simeq
 \frac{Z}{
   \omega
   -
   \Omega_L^{(0)}
   -
   gS_0
   +
   \ii\Gamma_L/2
 }
 +
 \frac{Z^{(\mathrm{drv})}}{
   \omega
   -
   \Omega_L^{(\mathrm{drv})}
   +
   \ii\Gamma_L^{(\mathrm{drv})}/2
 }.
 \label{eq:coeffCombinedSusceptibility}
\end{equation}
The drive-induced denominator contains no \(gS_0\) shift because the coherent \(b_+\) motion remains locked to the external drive frequency. In a coherently driven state, \(Z\) in Eq.~\eqref{eq:coeffCombinedSusceptibility} is the residue of the stationary fluctuation pole, whereas \(Z^{(\mathrm{drv})}\) is the residue of the drive-induced pole.

\section{Numerical validation from microscopic dynamics}
\label{app:numerical-validation}

To test the analytical frequency shifts, linewidths, and torque coefficients, we solve the semiclassical spin--phonon dynamics directly. The parameters used in the numerical calculations are summarized in Table~\ref{tab:suppValidationParameters}.

\begin{table*}[t]
\caption{Parameters used in the numerical calculations.}
\label{tab:suppValidationParameters}
\begin{ruledtabular}
\begin{tabular}{lclc}
Quantity & Value & Quantity & Value \\
\hline
Spin angular momentum \(S_0\) & \(1\hbar\) per unit cell &
Spin frequency \(\omega_s/(2\pi)\) & \(1.95\) THz \\
Gilbert damping \(\alpha_0\) & \(10^{-3}\) &
Land\'e factor \(g_{\mathrm L}\) & \(2\) \\
In-plane phonon frequency \(\omega_{\parallel}/(2\pi)\) & \(10\) THz &
Out-of-plane phonon frequency \(\omega_{\perp}/(2\pi)\) & \(12\) THz \\
In-plane phonon linewidth \(\kappa_{\parallel}/(2\pi)\) & \(2\) GHz &
Out-of-plane phonon linewidth \(\kappa_{\perp}/(2\pi)\) & \(10\) GHz \\
PAM coherence linewidth \(\Gamma_L/(2\pi)\) & \(12\) GHz &
Longitudinal PAM \(L_0\) & \(0.125\hbar\) per unit cell \\
PAM residue \(Z\) & \(0.252\hbar\) per unit cell &
Mode effective charge \(Z_{\rm mode}^\ast\) & \(1\,e/\sqrt{\mathrm{amu}}\) \\
Electric-pulse amplitude & \(0.15\) MV/cm &
Electric-pulse frequency \(\omega_E/(2\pi)\) & \(10\) THz \\
Electric-pulse center & \(2.5\) ps &
Electric-pulse width \(\sigma_E\) & \(0.40\) ps \\
Electric-pulse helicity & \(+1\) &
Magnetic-pulse amplitude & \(0.50\) T \\
Magnetic-pulse frequency \(\omega_B/(2\pi)\) & \(1.95\) THz &
Magnetic-pulse center & \(8\) ps \\
Magnetic-pulse width \(\sigma_B\) & \(0.75\) ps &
Magnetic-pulse helicity & \(-1\) \\
Main-text spin--PAM coupling \(g/(2\pi)\) & \(-65.2\) GHz/\(\hbar\) &
Electron gyromagnetic ratio \(\gamma_{\mathrm{el}}/(2\pi)\) & \(-28.0\) GHz/T \\
\end{tabular}
\end{ruledtabular}
\end{table*}

The microscopic equations are
\begin{equation}
 \begin{aligned}
 \dot{\vb S}
 &=
 \frac{1}{1+\alpha_0^2}
 \left[
   \vb S\times\vb W(t)
   -
   \frac{\alpha_0}{S_0}
   \vb S\times
   \left(
     \vb S\times\vb W(t)
   \right)
 \right],
 \\
 \dot{\vb Q}
 &=
 \vb P
 -
 g\,\vb S\times\vb Q
 -
 \frac{1}{2}\boldsymbol\kappa\,\vb Q,
 \\
 \dot{\vb P}
 &=
 -\mathbf K\vb Q
 -
 g\,\vb S\times\vb P
 -
 \frac{1}{2}\boldsymbol\kappa\,\vb P,
 \end{aligned}
 \label{eq:numericalEquations}
\end{equation}
where
\begin{equation}
 \vb W(t)
 =
 \omega_s\uv z
 +
 g\,\vb L_{\mathrm{ph}}(t)
 +
 \vb h_\perp(t),
 \qquad
 \vb L_{\mathrm{ph}}(t)
 =
 \vb Q(t)\times\vb P(t),
 \label{eq:numericalEffectiveField}
\end{equation}
\(\mathbf K=\operatorname{diag}(\omega_{\parallel}^2,\omega_{\parallel}^2,\omega_{\perp}^2)\), and \(\boldsymbol\kappa=\operatorname{diag}(\kappa_{\parallel},\kappa_{\parallel},\kappa_{\perp})\). The initial in-plane phonon motion is chosen circular, \(Q_x(0)=Q_0\), \(P_y(0)=\omega_{\parallel}Q_0\), with the remaining in-plane components zero, so that \(L_0=\omega_{\parallel}Q_0^2\). For calculations about a stationary circular background, we set \(\kappa_{\parallel}=0\) and \(\kappa_{\perp}=\Gamma_L\). This keeps the circular in-plane motion stationary while preserving the coherence linewidth \(\Gamma_L=\kappa_{\parallel}+\kappa_{\perp}\).

\begin{figure*}[t]
\centering
\includegraphics[width=\textwidth]{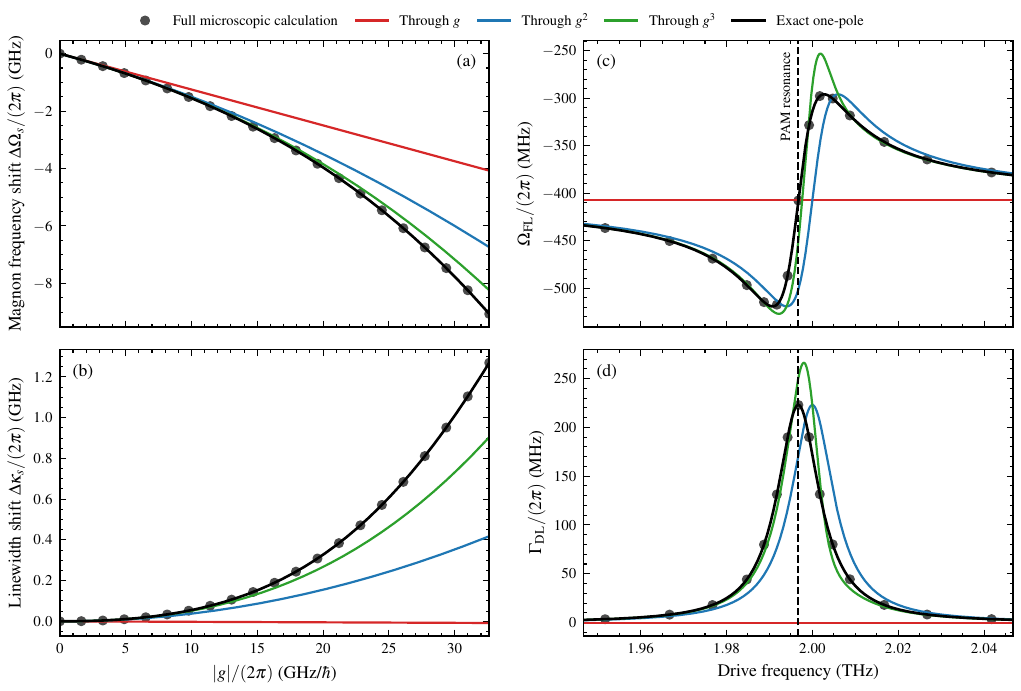}
\caption{
Numerical validation of the spin--PAM coupling.
Gray circles show the full microscopic calculation, the red, blue, and green curves show the perturbative results through order \(g\), \(g^2\), and \(g^3\), respectively, and the black curves show the exact one-pole results.
(a) Magnon frequency shift \(\Delta\Omega_s\) and (b) linewidth shift \(\Delta\kappa_s\) as \(\abs{g}/(2\pi)\) is varied from \(0\) to \(32.6\) GHz/\(\hbar\).
(c) Field-like torque coefficient \(\Omega_{\mathrm{FL}}\) and (d) damping-like torque coefficient \(\Gamma_{\mathrm{DL}}\) versus drive frequency for \(\abs{g}/(2\pi)=3.26\) GHz/\(\hbar\). The black dashed line marks the shifted PAM resonance.
}
\label{fig:suppCouplingValidation}
\end{figure*}

For Fig.~\ref{fig:suppCouplingValidation}(a),(b), we follow the spin-dominated complex eigenfrequency of the full microscopic dynamics as the coupling is varied. Its frequency and linewidth shifts are compared with the perturbative results in Eqs.~\eqref{eq:microFrequencyExpansion} and \eqref{eq:microLinewidthExpansion} and with the exact spin-dominated root of Eq.~\eqref{eq:microPoleEquation}.

For Fig.~\ref{fig:suppCouplingValidation}(c),(d), we use the weak circular transverse drive
\begin{equation}
 \vb h_\perp(t)
 =
 h_{\mathrm d}
 \left[
   \cos(\omega t)\uv x
   -
   \sin(\omega t)\uv y
 \right].
 \label{eq:numericalCircularDrive}
\end{equation}
The interaction contribution to the spin torque is
\begin{equation}
 \vb\tau_{\mathrm{ph}}
 =
 \frac{g}{1+\alpha_0^2}
 \left[
   \vb S\times\vb L_{\mathrm{ph}}
   -
   \frac{\alpha_0}{S_0}
   \vb S\times
   \left(
     \vb S\times\vb L_{\mathrm{ph}}
   \right)
 \right].
 \label{eq:numericalMicroscopicTorque}
\end{equation}
The complex amplitudes \(S_+(\omega)\) and \(\tau_{\mathrm{ph},+}(\omega)\) are obtained from the linear steady-state response of the microscopic equations. Using the torque decomposition in Eq.~\eqref{eq:coeffTorqueDecomposition},
\begin{equation}
 \frac{
   \tau_{\mathrm{ph},+}(\omega)
 }{
   S_+(\omega)
 }
 =
 -\Gamma_{\mathrm{DL}}(\omega)
 -
 \ii\Omega_{\mathrm{FL}}(\omega),
 \label{eq:numericalTorqueExtraction}
\end{equation}
so that
\begin{equation}
 \Omega_{\mathrm{FL}}(\omega)
 =
 -\operatorname{Im}
 \frac{\tau_{\mathrm{ph},+}(\omega)}{S_+(\omega)},
 \qquad
 \Gamma_{\mathrm{DL}}(\omega)
 =
 -\operatorname{Re}
 \frac{\tau_{\mathrm{ph},+}(\omega)}{S_+(\omega)}.
 \label{eq:numericalTorqueCoefficients}
\end{equation}
The resulting torque spectra are compared with the exact shifted one-pole expressions in Eqs.~\eqref{eq:coeffFieldLikeExact} and \eqref{eq:coeffDampingLikeExact} and with the perturbative expansions in Eqs.~\eqref{eq:coeffFieldLikeExpansion} and \eqref{eq:coeffDampingLikeExpansion}.

For the main-text pulse simulation, we use Eqs.~\eqref{eq:numericalEquations} and \eqref{eq:numericalEffectiveField}. We first prepare the longitudinal phonon angular momentum with a circular electric-field pulse that drives the degenerate in-plane phonons,
\[
\vb E(t)=E_0 e^{-(t-t_E)^2/(2\sigma_E^2)}
\left[\cos\!\bigl(\omega_E(t-t_E)\bigr)\uv x
+\eta_E\sin\!\bigl(\omega_E(t-t_E)\bigr)\uv y\right],
\]
which couples to the phonon mode effective charge \(Z_{\rm mode}^\ast\) and generates a driving force \(\vb F_E(t)=Z_{\rm mode}^\ast\vb E(t)\) that is added to \(\dot{\vb P}\) in Eq.~\eqref{eq:numericalEquations}. Here \(E_0\) is the peak electric-field amplitude, \(\omega_E\) is the carrier angular frequency, \(t_E\) is the pulse center, \(\sigma_E\) is the standard deviation of the Gaussian field-amplitude envelope, and \(\eta_E\) sets the helicity. We then excite the spin with a delayed circular magnetic-field pulse,
\[
\vb B(t)=B_0 e^{-(t-t_B)^2/(2\sigma_B^2)}
\left[\cos\!\bigl(\omega_B(t-t_B)\bigr)\uv x
+\eta_B\sin\!\bigl(\omega_B(t-t_B)\bigr)\uv y\right],
\]
which enters Eq.~\eqref{eq:numericalEffectiveField} through \(\vb h_\perp(t)=\gamma_{\mathrm{el}}\vb B(t)\). Here \(B_0\), \(\omega_B\), \(t_B\), and \(\sigma_B\) are the corresponding magnetic-pulse amplitude, carrier angular frequency, pulse center, and Gaussian-envelope standard deviation. The pulse delay is \(t_B-t_E\). We use \(\eta_E=+1\) and \(\eta_B=-1\).

\section{Microscopic dynamical simulations, exchange, and hybrid spin--PAM modes}
\label{app:microscopic-dynamical-simulations}

We complement the response-function calculation with direct semiclassical simulations of the microscopic spin and phonon variables. Unless otherwise stated, the representative calculations in this section use the same microscopic dynamics as Eqs.~\eqref{eq:numericalEquations} and \eqref{eq:numericalEffectiveField} with the stationary-background damping choice \(\kappa_{\parallel}=0\) and \(\kappa_{\perp}=\Gamma_L\); where needed, we allow the coupling to vary in time, \(g\to g(t)\). We calculate the phonon angular momentum at every time step from
\begin{equation}
 \vb L_{\mathrm{ph}}(t)=\vb Q(t)\times\vb P(t).
 \label{eq:numInstantaneousLph}
\end{equation}
The spin equation is
\begin{equation}
 \dot{\vb S}
 =
 \frac{1}{1+\alpha_0^2}
 \left[
   \vb S\times\vb W(t)
   -
   \frac{\alpha_0}{S_0}
   \vb S\times
   \left(
     \vb S\times\vb W(t)
   \right)
 \right],
 \qquad
 \vb W(t)
 =
 \omega_s\uv z
 +
 g(t)\vb L_{\mathrm{ph}}(t)
 +
 \vb h_{\mathrm d}(t),
 \label{eq:numSpinEquation}
\end{equation}
where \(\vb h_{\mathrm d}(t)\) is included only in the driven calculations. The phonon variables obey
\begin{equation}
 \dot{\vb Q}
 =
 \vb P
 -
 g(t)\vb S\times\vb Q
 -
 \frac{\Gamma_L}{2}Q_z\uv z,
 \qquad
 \dot{\vb P}
 =
 -\mathbf K\vb Q
 -
 g(t)\vb S\times\vb P
 -
 \frac{\Gamma_L}{2}P_z\uv z.
 \label{eq:numPhononEquations}
\end{equation}
The last terms implement this damping choice. The in-plane circular phonon is undamped, while the out-of-plane amplitude decays with rate \(\Gamma_L/2\), so the transverse phonon-angular-momentum coherence has full linewidth \(\Gamma_L\). In the conservative limit, Eq.~\eqref{eq:numPhononEquations} reduces to Eq.~\eqref{eq:microInteractingHamiltonEquations}, while Eq.~\eqref{eq:numSpinEquation} reduces to the undamped spin dynamics with the interaction torque in Eq.~\eqref{eq:microSpinTorque}.

\subsection{Coupling switch-on and emergence of transverse phonon angular momentum}

\begin{figure*}[t]
 \centering
 \includegraphics[width=\textwidth]{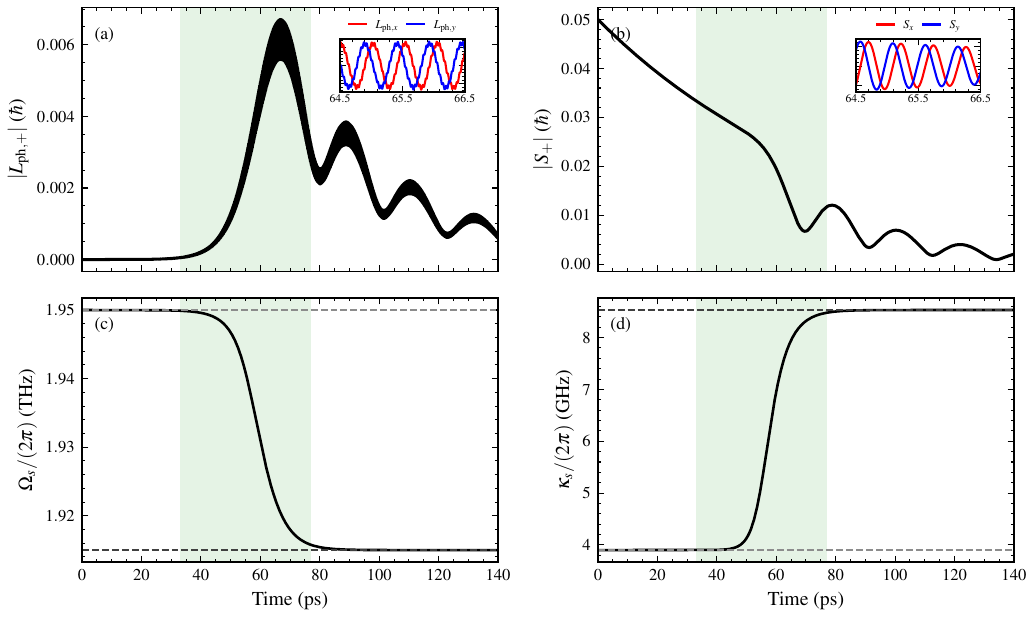}
 \caption{Semiclassical dynamics when the spin--phonon-angular-momentum coupling is switched on. The green region marks the switching interval. (a) Transverse PAM amplitude \(\abs{L_{\mathrm{ph},+}}=\sqrt{L_{\mathrm{ph},x}^2+L_{\mathrm{ph},y}^2}\); the inset resolves the \(x\) and \(y\) components. (b) Transverse spin amplitude \(\abs{S_+}=\sqrt{S_x^2+S_y^2}\); the inset resolves \(S_x\) and \(S_y\). (c),(d) Spin-dominated resonance frequency \(\Omega_s\) and linewidth \(\kappa_s\) evaluated at the instantaneous value of \(g(t)\). The horizontal dashed lines indicate the uncoupled and final coupled values.}
 \label{fig:numSwitchOnDynamics}
\end{figure*}

We initialize the lattice in a circular in-plane orbit with \(Q_x(0)=Q_0\), \(P_y(0)=\omega_{\parallel} Q_0\), and all other phonon components zero. This gives \(L_{\mathrm{ph},z}(0)=L_0=\omega_{\parallel} Q_0^2\). The spin is given a small transverse tilt so that it undergoes a free ringdown. The coupling is initially zero and is then switched smoothly to a finite value. Before the switch, the spin and phonons evolve independently and \(L_{\mathrm{ph},x}=L_{\mathrm{ph},y}=0\). Once the coupling is active, the transverse spin motion drives \(Q_z\) and \(P_z\), and the cross products between the original in-plane circular motion and the induced out-of-plane motion generate oscillating \(L_{\mathrm{ph},x}\) and \(L_{\mathrm{ph},y}\). The phonon angular-momentum vector therefore acquires a transverse precessing component, as shown in Fig.~\ref{fig:numSwitchOnDynamics}.

The same switch changes the spin-dominated resonance frequency and linewidth from their uncoupled to coupled values. Fig.~\ref{fig:numSwitchOnDynamics}(c),(d) shows the spin-dominated complex eigenfrequency evaluated at the instantaneous value of \(g(t)\).

\subsection{Angular-momentum and energy exchange}

\begin{figure*}[t]
 \centering
 \includegraphics[width=\textwidth]{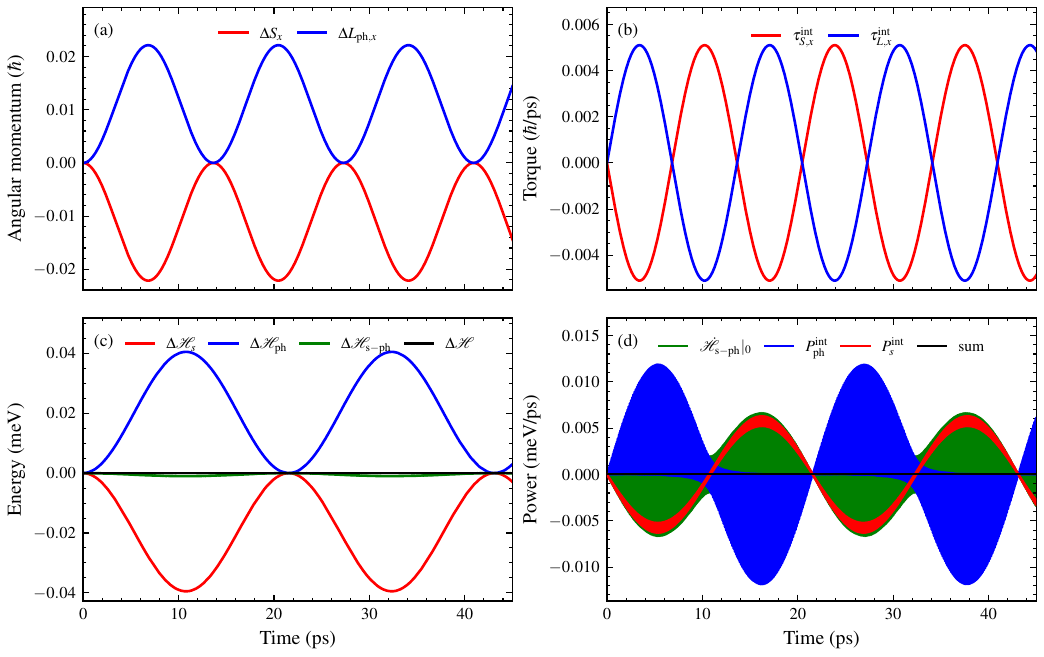}
 \caption{Angular-momentum and energy exchange in the microscopic dynamics. (a) Changes \(\Delta S_x\) and \(\Delta L_{\mathrm{ph},x}\) in a conservative isotropic control calculation. Their opposite signs show direct angular-momentum transfer. (b) Interaction torques \(\tau_{S,x}^{\mathrm{int}}\) and \(\tau_{L,x}^{\mathrm{int}}\), which are equal and opposite at every time. (c) Changes of the spin, phonon, interaction, and total Hamiltonians in a conservative anisotropic calculation. The total energy remains constant while the subsystem energies exchange periodically. (d) Interaction powers \(P_s^{\mathrm{int}}\), \(P_{\mathrm{ph}}^{\mathrm{int}}\), and \(\left.\dot{\mathcal H}_{\mathrm{s-ph}}\right|_0\); their sum remains zero to numerical accuracy.}
 \label{fig:numExchangeSummary}
\end{figure*}

The interaction torque is equal and opposite in the two angular-momentum sectors,
\begin{equation}
 \vb\tau_S^{\mathrm{int}}
 =
 \left.\dot{\vb S}\right|_{\mathrm{s-ph}}
 =
 g\,\vb S\times\vb L_{\mathrm{ph}},
 \qquad
 \vb\tau_L^{\mathrm{int}}
 =
 \left.\dot{\vb L}_{\mathrm{ph}}\right|_{\mathrm{s-ph}}
 =
 -g\,\vb S\times\vb L_{\mathrm{ph}}.
 \label{eq:numExchangeTorques}
\end{equation}
To isolate this exchange from external-field, anisotropy, and damping torques, we perform a conservative isotropic control calculation. In that limit,
\begin{equation}
 \frac{\dd}{\dd t}
 \left(
   \vb S+\vb L_{\mathrm{ph}}
 \right)
 =
 0,
 \label{eq:numTotalAngularMomentumConservation}
\end{equation}
and the simulations give \(\Delta\vb S(t)=-\Delta\vb L_{\mathrm{ph}}(t)\) together with \(\vb\tau_S^{\mathrm{int}}(t)=-\vb\tau_L^{\mathrm{int}}(t)\), as shown in Fig.~\ref{fig:numExchangeSummary}(a),(b). In the full anisotropic and damped calculation, additional field, anisotropy, and reservoir torques act on the system, so \(\vb S+\vb L_{\mathrm{ph}}\) is not generally conserved. The interaction contributions in Eq.~\eqref{eq:numExchangeTorques}, however, remain equal and opposite.

We also resolve the corresponding energy flow. For the conservative calculation, we separate
\begin{equation}
 \mathcal H
 =
 \mathcal H_s
 +
 \mathcal H_{\mathrm{ph}}
 +
 \mathcal H_{\mathrm{s-ph}},
 \qquad
 \mathcal H_s=-\omega_sS_z,
 \qquad
 \mathcal H_{\mathrm{s-ph}}
 =
 -g\,\vb S\cdot
 \left(
   \vb Q\times\vb P
 \right),
 \label{eq:numEnergyDecomposition}
\end{equation}
with \(\mathcal H_{\mathrm{ph}}\) given by Eq.~\eqref{eq:microPhononHamiltonian}. The powers delivered by the interaction to the spin and phonon sectors are
\begin{equation}
 P_s^{\mathrm{int}}
 =
 \frac{\partial\mathcal H_s}{\partial\vb S}
 \cdot
 \left.\dot{\vb S}\right|_{\mathrm{s-ph}},
 \qquad
 P_{\mathrm{ph}}^{\mathrm{int}}
 =
 \frac{\partial\mathcal H_{\mathrm{ph}}}{\partial\vb Q}
 \cdot
 \left.\dot{\vb Q}\right|_{\mathrm{s-ph}}
 +
 \frac{\partial\mathcal H_{\mathrm{ph}}}{\partial\vb P}
 \cdot
 \left.\dot{\vb P}\right|_{\mathrm{s-ph}}.
 \label{eq:numInteractionPowers}
\end{equation}
The interaction energy also changes under the uncoupled spin and phonon dynamics. We denote this contribution by
\begin{equation}
 \left.\dot{\mathcal H}_{\mathrm{s-ph}}\right|_0
 =
 \frac{\partial\mathcal H_{\mathrm{s-ph}}}{\partial\vb S}
 \cdot
 \left.\dot{\vb S}\right|_0
 +
 \frac{\partial\mathcal H_{\mathrm{s-ph}}}{\partial\vb Q}
 \cdot
 \left.\dot{\vb Q}\right|_0
 +
 \frac{\partial\mathcal H_{\mathrm{s-ph}}}{\partial\vb P}
 \cdot
 \left.\dot{\vb P}\right|_0.
 \label{eq:numInteractionEnergyRate}
\end{equation}
Hamiltonian energy conservation then requires
\begin{equation}
 P_s^{\mathrm{int}}
 +
 P_{\mathrm{ph}}^{\mathrm{int}}
 +
 \left.\dot{\mathcal H}_{\mathrm{s-ph}}\right|_0
 =
 0.
 \label{eq:numPowerBalance}
\end{equation}
Figs.~\ref{fig:numExchangeSummary}(c) and \ref{fig:numExchangeSummary}(d) show the periodic exchange among the spin, phonon, and interaction energies and verify Eq.~\eqref{eq:numPowerBalance}. Thus, the coupling both shifts the spin resonance and transfers angular momentum and energy between the magnetic and lattice sectors.

\subsection{Collective PAM mode and hybridization with the magnon}

The transverse phonon-angular-momentum response is a coherence between the in-plane circular phonon and the out-of-plane phonon. Its uncoupled frequency is the difference frequency \(\Omega_L^{(0)}=\omega_{\perp}-\omega_{\parallel}\), shifted to \(\Omega_L^{(0)}+gS_0\) by the longitudinal spin polarization. It is therefore a collective mode of the phonon angular momentum. The reduced linear equations, Eqs.~\eqref{eq:microCoupledSpin} and \eqref{eq:microCoupledL}, describe its hybridization with the magnon.

The determinant of Eq.~\eqref{eq:microCoupledMatrix} gives
\begin{equation}
 \left[
   \omega
   -
   \frac{\omega_s+gL_0}{1+\ii\alpha_0}
 \right]
 \left[
   \omega
   -
   \Omega_L^{(0)}
   -
   gS_0
   +
   \ii\Gamma_L/2
 \right]
 =
 \frac{g^2S_0Z}{2(1+\ii\alpha_0)}.
 \label{eq:numTwoModePoleEquation}
\end{equation}
The two complex eigenfrequencies are consequently
\begin{equation}
 \begin{aligned}
 \widetilde\Omega_{\pm}
 &=
 \frac{1}{2}
 \left[
   \frac{\omega_s+gL_0}{1+\ii\alpha_0}
   +
   \Omega_L^{(0)}
   +
   gS_0
   -
   \ii\Gamma_L/2
 \right]
 \\
 &\quad
 \pm
 \sqrt{
   \frac{1}{4}
   \left[
     \frac{\omega_s+gL_0}{1+\ii\alpha_0}
     -
     \Omega_L^{(0)}
     -
     gS_0
     +
     \ii\Gamma_L/2
   \right]^2
   +
   \frac{g^2S_0Z}{2(1+\ii\alpha_0)}
 }.
 \end{aligned}
 \label{eq:numHybridPoles}
\end{equation}
Their resonance frequencies and linewidths are
\begin{equation}
 \Omega_\pm
 =
 \operatorname{Re}
 \widetilde\Omega_\pm,
 \qquad
 \kappa_\pm
 =
 -2\operatorname{Im}
 \widetilde\Omega_\pm.
 \label{eq:numHybridFrequencyLinewidth}
\end{equation}
We next obtain the normal modes of the coupled spin and lattice dynamics in the absence of an external drive. We linearize the coupled \((\vb S,\vb Q,\vb P)\) equations, Eqs.~\eqref{eq:numSpinEquation} and \eqref{eq:numPhononEquations}, about the circular in-plane phonon background carrying the prescribed longitudinal PAM \(L_0\uv z\), with the spin polarized along \(S_0\uv z\). This background rotates in the laboratory frame at \(\omega_{\parallel}-gS_0\), so we transform to the frame co-rotating with the in-plane phonon, where it is stationary and the linearized coefficients are time independent. Writing \(\delta\vb X=(\delta\vb S,\delta\vb Q,\delta\vb P)^T\), the linearized dynamics have the form
\begin{equation}
 \delta\dot{\vb X}
 =
 \mathbf M\,\delta\vb X,
 \qquad
 \mathbf M
 =
 \left.
 \frac{\partial\vb F_{\mathrm{rot}}}{\partial\vb X}
 \right|_{\vb X=\vb X_0},
 \label{eq:numMicroscopicDynamicalMatrix}
\end{equation}
where \(\mathbf M\) is the \(9\times9\) microscopic dynamical matrix, \(\vb F_{\mathrm{rot}}\) denotes the right-hand side of Eqs.~\eqref{eq:numSpinEquation} and \eqref{eq:numPhononEquations} in the co-rotating frame, and \(\vb X_0\) is the stationary circular background. After transformation back to the laboratory frame, the eigenvalues of \(\mathbf M\) give the complex normal-mode frequencies of the complete microscopic system. By contrast, Eq.~\eqref{eq:numHybridPoles} is the reduced two-mode result obtained by retaining only the magnon amplitude \(S_+\) and the resonant PAM amplitude \(\ell_+\).

\begin{figure*}[t]
 \centering
 \includegraphics[width=\textwidth]{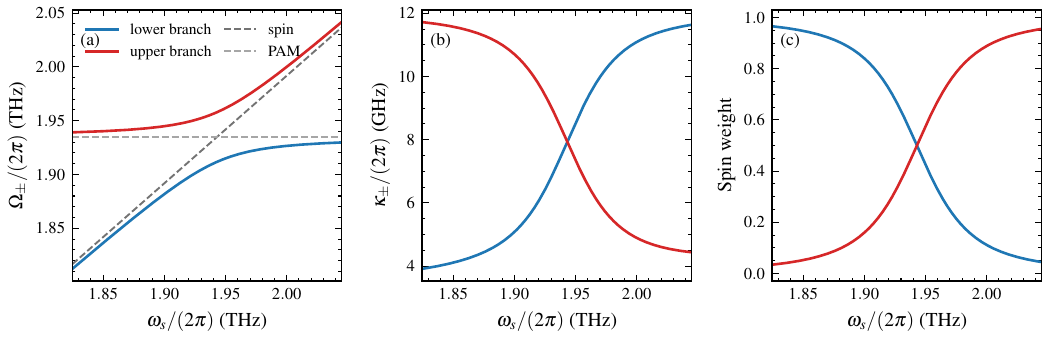}
 \caption{Hybridization of the magnon and the PAM difference-frequency mode. (a) Microscopic coupled-mode frequencies as the bare spin frequency \(\omega_s\) is varied through the PAM resonance. The dashed curves are the uncoupled spin and PAM reference branches obtained by setting the off-diagonal terms in Eq.~\eqref{eq:microCoupledMatrix} to zero while retaining the longitudinal shifts. The solid branches avoid crossing. (b) Linewidths \(\kappa_\pm\) of the same coupled modes. Their interchange reflects the exchange of spin and PAM character. (c) Spin weight defined in Eq.~\eqref{eq:numSpinWeight}. At the center of the avoided crossing both modes are equally mixed, while far from resonance the branches recover predominantly spin-like or PAM-like character.}
 \label{fig:numHybridization}
\end{figure*}

Fig.~\ref{fig:numHybridization}(a) shows the resonance frequencies of the two positive-frequency microscopic modes that evolve from the magnon and PAM difference-frequency modes as the bare spin frequency \(\omega_s\) is varied through resonance. The avoided crossing originates from the mutual-conversion terms \(gS_0\) and \(gZ/2\) in Eq.~\eqref{eq:microCoupledMatrix}. Fig.~\ref{fig:numHybridization}(b) shows the linewidths of the same modes, with \(\kappa=-2\operatorname{Im}\widetilde\Omega\) as in Eq.~\eqref{eq:numHybridFrequencyLinewidth}. The spin weight in Fig.~\ref{fig:numHybridization}(c) is obtained from the same microscopic eigenvectors after projection onto \(S_+\) and \(\ell_+\):
\begin{equation}
 w_s
 =
 \frac{|S_+|^2/S_0}
 {|S_+|^2/S_0+|\ell_+|^2/(Z/2)}.
 \label{eq:numSpinWeight}
\end{equation}
Here \(w_s=1\) and \(0\) denote purely spin- and PAM-like character, respectively. Far from resonance the two branches are predominantly magnon-like and PAM-like, while near the center of the avoided crossing they are strongly mixed.

As an independent time-domain test, we apply a weak circular transverse drive to the spin,
\begin{equation}
 \vb h_{\mathrm d}(t)
 =
 h_{\mathrm d}
 \left[
   \cos(\omega t)\uv x
   -
   \sin(\omega t)\uv y
 \right]
 \label{eq:numCircularSpinDrive}
\end{equation}
and evolve the microscopic equations to the periodic steady state. The complex response amplitudes are extracted from
\begin{equation}
 S_+(\omega)
 =
 \avg{
   S_+(t)\ee^{+\ii\omega t}
 }_{\mathrm{ss}},
 \qquad
 L_{\mathrm{ph},+}(\omega)
 =
 \avg{
   L_{\mathrm{ph},+}(t)\ee^{+\ii\omega t}
 }_{\mathrm{ss}},
 \label{eq:numSteadyStatePhasors}
\end{equation}
where the average is taken after the transient has decayed. The resonances obtained from these driven responses follow the same hybrid branches \(\Omega_\pm\) shown in Fig.~\ref{fig:numHybridization}(a), providing an independent driven-dynamics signature of the same hybridization.

The microscopic simulations therefore establish that a precessing spin generates transverse phonon angular momentum, the retarded lattice response shifts the magnon frequency and linewidth, and the coupling transfers both angular momentum and energy between the magnetic and lattice sectors. The dynamical-matrix eigenvalues and eigenvectors identify the difference-frequency coherence encoded in \(\vb L_{\mathrm{ph}}=\vb Q\times\vb P\) as a collective PAM mode that hybridizes with the magnon. This mode is a collective coherence of the in-plane and out-of-plane phonons.

\end{document}